\documentclass[aps,  pra,  superscriptaddress,  showpacs]{revtex4-1}
\usepackage[T1]{fontenc}
\usepackage[utf8]{inputenc}
\usepackage[english]{babel}
\usepackage{amsmath,amssymb,amsfonts}
\usepackage{graphicx,graphics}
\usepackage{subfig}
 \usepackage{caption}
 \usepackage{dcolumn}
\usepackage{bm}
\usepackage{color}
\usepackage{amsthm}             
\usepackage{amssymb}
\usepackage{mathrsfs}
\usepackage{dcolumn}
\usepackage{bm}
\usepackage{epsfig}
\usepackage{epstopdf}
\usepackage{float}

\begin{document}

\title{On demand generation of propagation invariant photons with orbital angular momentum }
\author{Y. Jer\'onimo-Moreno,  R. J\'auregui}

\address{Instituto de F\'isica, Universidad Nacional Aut\'onoma de M\'exico,\\ Apartado Postal 20-364, M\'exico D. F. 01000, M\'exico}
\email{rocio@fisica.unam.mx}
\begin{abstract}
 We study the generation of propagation invariant photons with orbital angular momentum  by spontaneous parametric down conversion (SPDC) using a Bessel-Gauss pump beam.  The angular and conditional angular spectra are calculated for an uniaxial  crystal  optimized for type I SPDC with standard Gaussian pump beams. It is shown that, as the mean value of the magnitude of the transverse wave vector of the pump beam increases, the emission cone is deformed into two non coaxial cones that touch each other along a line determined by the orientation of the optical axis of the nonlinear crystal. At this location, the conditional spectrum becomes maximal for a pair of photons, one of which is best described  by a Gaussian-like photon with a very small transverse wave vector, and the other a Bessel-Gauss photon with a distribution of transverse wave vectors similar in amplitude to that of the incident pump beam. A detailed analysis is then performed of the angular momentum content of SPDC photons by the evaluation of the corresponding transition amplitudes. As a result, we obtain conditions for the generation of heralded single photons which are approximately propagation invariant and have orbital angular momentum. A discussion is given about the difficulties in the interpretation of the results in terms of conservation of optical orbital angular momentum along the vector normal to the crystal surface. The angular spectra and the conditional angular spectra are successfully compared with available experimental data recently reported in the literature.
\end{abstract}

\maketitle

\section{Introduction}

In the last two decades there has been huge advances in the generation of structured light beams with several spatial and dynamical features.  Probably the best known examples correspond to beams with orbital angular momentum (OAM), e. g. Laguerre \cite{allen} and Bessel beams \cite{dholakia}, though other possibilities with diverse transverse \cite{blas,julio} and longitudinal structure \cite{christo} are not less interesting. The richness of structured beams is inherited to several optical phenomena. Here we shall focus on spontaneous parametric down conversion (SPDC).

SPDC from structured beams can be useful for the implementation of quantum information protocols for which the encoding variables may be different from  polarization, linear momentum or frequency of the photons. In quantum optics, two-photon states entangled in polarization, usually generated via a SPDC process, have become the most widely used entangled systems. In this case,  the two photons in a pair usually have one of two orthogonal polarization directions \cite{Kwiat1995}. In most cases, the description of the system is made in a Hilbert space of just two-dimensions, though the role of the other  degrees of freedom of the photon should not be ignored in general. Quantum-enhanced technologies demand systems consisting of multiple entangled states, as well as quantum states entangled in multiple dimensions. The latter may, for instance, improve the communication channel efficiency in quantum cryptography \cite{Bechmann2000}. The generation of two-photon states with multi-dimensional entanglement has been shown to be feasible, and the challenge to control it is the main subject of many theoretical and experimental current studies.

One way to produce entanglement in a Hilbert space with dimension greater than two, is by taking advantage of the diversity of spatial modes of the electromagnetic (EM) field.  Furthermore, from  modes with well defined OAM,  it is possible to generate two-photon states entangled in this degree of freedom, which have a discrete dimensionality as a result of the quantization of angular momentum. The first experimental demonstration of OAM entanglement in photon pairs generated by SPDC was reported in 2001 \cite{Mair2001} and since then several similar sources have been implemented.

The role of the linear and nonlinear electromagnetic response of the material used for SPDC on the pump beam is essential for a
precise description of the expected quantum correlations of the photon pairs. There is not a unique basis set to perform such a description. Most studies of SPDC rely on the use of the plane EM wave expansion. Accordingly, polarization, linear momentum and frequency become the natural degrees of freedom, and the effects of birefringence  on these variables through the type I and type II phase matching  conditions restricts the possibility of  SPDC by conventional nonlinear crystals. For pump beams with OAM, one may describe the system as a continuum superposition of plane EM waves, or perform the description in terms of modes with OAM. In the latter case it must be recognized that the birefringence of SPDC crystals is able to modify both the polarization and the OAM of the modes that propagate in them \cite{Hacyan2009}. As a consequence, OAM conservation in SPDC does not necessarily hold \cite{Osorio2008}. Most studies of this kind of effects take as starting point an effective scalar treatment of the EM field and the paraxial description of the pump mode, though going beyond these approximations may lead to observable effects \cite{monken2008,Osorio2008,fedorov2008}.

As already reported in Ref.~\cite{monken1}, for wide crystals, the plane-wave spectrum from the pump beam results
in a multiplicative factor for the transition amplitude of the generated two-photon state. This opens the possibility
of using SPDC to generate photons with general properties inherited from the pump photons. In the case of  a propagation invariant Bessel pump beam, this could lead to the generation of propagation invariant  photons with
OAM. In the present work, we theoretically study the circumstances under which this process is feasible and compare our analysis
with available experimental results \cite{uren2012}. Notice that in the implementation of many quantum protocols, the photons
with selected properties are generated in a given space region and are processed in a different location. Propagation
invariant photons have an angular spectra restricted to a cone in wave vector space. Its perpendicular radius can be optimized for an efficient coupling to optical fibers \cite{sergienko}. Besides, they possess the self healing property \cite{bouchal}, that is they tend to reform during propagation in spite of blocking part of them. This property makes them robust in scattering and turbulent environments.

The article is organized as follows. In section II, we describe  general features of  type I SPDC  involving structured vectorial beams in a birefringent medium. In section III, this formalism is applied to the case of a Bessel-Gauss pump beam. Explicit analytical expressions and numerical results for the angular spectrum are given and compared with experimental results. The angular momentum properties of the photon pair are also analyzed, including a discussion on the pertinence of studying its conservation in the SPDC process. Finally,  we outline conclusions derived from this study.

\section{SPDC of structured EM fields.}

Spontaneous parametric down conversion focus on the evolution of an initial state of the quantum electromagnetic field that corresponds to a coherent state for a given pumping mode $\kappa_p$ with no other occupied mode $\vert 0; \alpha_{\kappa_p}\rangle$, under the $\hat{\mathcal{H}}_{SPDC}$ Hamiltonian \cite{HongMandel},
\begin{equation}
\hat{\mathcal{H}}_{SPDC} \vert 0; \alpha_{\kappa_p}\rangle = \frac{1}{2} \sum_{r,l,t}\int d^3r \chi_{r,l,t}^{(2)}\hat{\mathbf{E}}_r^{(+)} \hat{\mathbf{E}}_l^{(+)} \hat{\mathbf{E}_t}^{(-)}\vert 0; \alpha_{\kappa_p}\rangle .
\end{equation}
Here  $\chi_{r,l,t}^{(2)}$ denotes the second order electric susceptibility of the nonlinear media. The operators $\hat{\mathbf{E}}_l^{(\pm)}$ are usually written as a series expansion on a basis formed by EM  monochromatic modes that
satisfy the adequate boundary conditions. $\hat{\mathbf{E}}_l^{(+)}$ ($\hat{\mathbf{E}}_l^{(-)}$) is the electric field operator that includes the
modes with frequency $\omega$ that propagate with a time dependence $e^{-i\omega t}$ ($e^{i\omega t}$ ). The best known example of $\hat{\mathbf{E}}$ corresponds to the EM field  confined within a rectangular cavity of volume $V$ in otherwise free space:
\begin{eqnarray}
 \hat{\mathbf{E}} &=&  \hat{\mathbf{E}}^{(+)} +\hat{\mathbf{E}}^{(-)}\nonumber\\
 & = & \sum_s\int d\omega d^3\mathbf{k} \mathcal{N}_{\mathbf{k}} \delta(\sqrt{k_x^2 +k_y^2 +k_z^2} -\omega/c)
 \Big[ e^{i(\mathbf{k}\cdot\mathbf{r} - \omega t)} \hat{e}_{\mathbf{k},s} \hat{a}_{\mathbf{k},s}^\dagger
     + e^{-i(\mathbf{k}\cdot\mathbf{r} - \omega t)} \hat{e}_{\mathbf{k},s}^*\hat{a}_{\mathbf{k},s}\Big],
\end{eqnarray}
$\hat{e}_{\mathbf{k},s}$ denotes the unitary vectors giving the mode polarization $s$, $c$  is the velocity of light in vacuum and $\mathcal{N}_{\mathbf{k}}$
is the normalization factor, $\mathcal{N}_{\mathbf{k}}= \sqrt{2\pi\hbar/\omega_{\mathbf{k}} V}$. For other boundary conditions or in the presence of a media that modifies the dispersion relations, each electromagnetic mode $\kappa$ can still be expressed in terms of its properly normalized  plane-wave spectrum ${\mathfrak{E}}_{\kappa,l}^{(+)}(s,\mbox{k}_x,\mbox{k}_y;\omega)$. In this case,
\begin{eqnarray}
\hat{\mathcal{H}}_{SPDC} \vert 0; \alpha_{\kappa_p}\rangle &=& \sum_{\kappa^s,\kappa^i}\Big[\frac{1}{2}\sum_{r,l,t}\int d^6 \mbox{k}_\bot
\chi_{r,l,t}^{(2)}{\mathfrak{E}}_{\kappa^s,r}^{(+)}(s^s,\mbox{k}_x^s,\mbox{k}_y^s;\omega^s)
{\mathfrak{E}}_{\kappa^i,l}^{(+)}(s^i,\mbox{k}_x^i,\mbox{k}_y^i;\omega^i)
{\mathfrak{E}}_{\kappa^p,t}^{(-)}(s^p,\mbox{k}_x^p ,\mbox{k}_y^p;\omega^p)\nonumber\\
&\times &\int d^3r   e^{i(\Delta \mathbf{k}\cdot \mathbf{r} -\Delta \omega t)}\hat{a}^\dagger(s^s,\mbox{k}_x^s,\mbox{k}_y^s,\omega^s)\hat{a}^\dagger(s^i,\mbox{k}_x^i,\mbox{k}_y^i,\omega^i)\alpha_{\kappa^p}\vert 0; \alpha_{\kappa^p}\rangle\Big].\label{e:f}
\end{eqnarray}
Here $d^6k_{\bot}$ encloses the transverse momentum differentials for each field; $\Delta \mathbf{k}$ denotes the vectorial mismatch term $\Delta \mathbf{k} =\mathbf{k}^p - \mathbf{k}^s -\mathbf{k}^i$, and $\Delta\omega = \omega^p-\omega^s-\omega^i$. Each $\{\kappa^s,\kappa^i\}$ term in this expansion determines the probability amplitude for the creation of
a photon in the signal (idler) mode $\kappa^s$ ($\kappa^i$) together with the annihilation of a photon in the coherent pumping state.
The time integral of this equation gives an approximate expression, valid within first order perturbation theory, of the
time evolved SPDC state
\begin{eqnarray}
\label{E:St}
\vert \Psi(t)\rangle &=&  \vert 0; \alpha_{\kappa_p}\rangle +
\sum_{\kappa^s,\kappa^i}\Big[\frac{1}{2}\sum_{r,l,t}\int d^6k_\bot
\chi_{r,l,t}^{(2)}{\mathfrak{E}}_{\kappa^s,r}^{(+)}(s^s,\mbox{k}_x^s,\mbox{k}_y^s;\omega^s)
{\mathfrak{E}}_{\kappa^i,l}^{(+)}(s^i,\mbox{k}_x^i,\mbox{k}_y^i;\omega^i)
{\mathfrak{E}}_{\kappa^p,t}^{(-)}(s^p,\mbox{k}_x^p ,\mbox{k}_y^p;\omega^p) \nonumber\\
&\times &\int d^3r   \mbox{e}^{i(\Delta\mathbf{ k}\cdot \mathbf{r}-\Delta \omega t/2)}\mbox{sinc} (\Delta \omega t/2)\hat{a}^\dagger(s^s,\mbox{k}_x^s,\mbox{k}_y^s,\omega^s)\hat{a}^\dagger(s^i,\mbox{k}_x^i,\mbox{k}_y^i,\omega^i)\alpha_{\kappa^p}\vert 0; \alpha_{\kappa^p}\rangle\Big].
\end{eqnarray}

In this work we shall present an analysis for a type-I SPDC, with this configuration, for a quasi monochromatic pump beam  that propagates in an uniaxial crystal
with symmetry axis $\mathbf{a}$ and  permeability  coefficients $\epsilon_\|$ and $\epsilon_\bot$ parallel  and transversal   to the optical axis respectively. The electric field associated to the extraordinary waves is:
\begin{equation}
\label{eq:bmodeE}
{\mathfrak{E}}_{\mbox{e,}\kappa}(\mbox{k}_x ,\mbox{k}_y;\omega)= \left[- \frac{c^2}{\epsilon_\bot\omega^2}\mathbf{k}^{\mbox{e}}({\bf a} \cdot \mathbf{k}^{\mbox{e} })+ {\bf a}\right]
\mathcal{N}_{\mbox{e}}\alpha(\omega)\tilde{\psi}_\kappa(\mbox{k}_x ,\mbox{k}_y),
\end{equation}
with $\tilde{\psi}_\kappa(\mbox{k}_x ,\mbox{k}_y)$ the 2-dimensional Fourier transform of the pump beam evaluated at the
crystal surface and $\alpha(\omega)$ its spectral envelop.
As for the generated photons,
\begin{equation}
\label{eq:bmodeOO}
{\mathfrak{E}}_{{\mbox{o}},\mbox{k}_x^{0},\mbox{k}_y^{0}}(\mbox{k}_x,\mbox{k}_y;\omega^0)= {\bf a} \times {\bf k}^{\mbox{o}}\mathcal{N}_{\mbox{o}}\delta (\mbox{k}_x - \mbox{k}_x^{0})
\delta (\mbox{k}_y - \mbox{k}_y^{0}),
\end{equation}
 if they are described by  ordinary vectorial plane waves with wave vector $\mathbf{k}^{0}$ and frequency $\omega^0$.
In Eqs.~(\ref{eq:bmodeE}-\ref{eq:bmodeOO})  $\mathcal{N}_{\mbox{e}}$ and  $\mathcal{N}_{\mbox{o}}$ are the normalization factors, and the vectors $\mathbf{k}^{\mbox{e}}$ and $\mathbf{k}^{\mbox{o}}$ satisfy the extraordinary and the ordinary dispersion relations, respectively. Notice that, as expected the electric field of the ordinary modes is perpendicular to both the optical axis $\mathbf{a}$ and the wave vector $\mathbf{k}$, while the electric field of the extraordinary modes has a component along the optical axis $\mathbf{a}$ and a component along the wave vector $\mathbf{k}$ in a combination that guarantees the fulfillment of the Maxwell equations in the birefringent media.

For a wide crystal the state of the electromagnetic field at asymptotic times can be written as
\begin{equation}
\label{E:St}
\vert \Psi\rangle=  \vert 0; \alpha_{\kappa_p}\rangle
+ \frac{\pi}{i\hbar} \int d\omega^s\int d\omega^i\int d^2 \mbox{k}_\bot^{s} d^2 \mbox{k}_\bot^{i}\mathcal{N}_p\mathcal{N}_s\mathcal{N}_i
\alpha(\omega^s + \omega^i){\mathcal{X}}
F(\mbox{k}_x^s,\mbox{k}_y^s;\omega^s,\mbox{k}_x^i,\mbox{k}_y^i;\omega^i)\vert 0; \alpha_{\kappa_p};1_s;1_i\rangle.
\end{equation}
The factor ${\mathcal{X}}$  results from the contraction of the nonlinear susceptibility tensor $\chi_{r,l,q}^{(2)}$ with the polarization vectors,
while the  wave vector joint amplitude  is defined as
\begin{equation}
F(\mbox{k}_x^s,\mbox{k}_y^s;\omega^s,\mbox{k}_x^i,\mbox{k}_y^i;\omega^i)= \tilde{\psi}_\kappa
(\mbox{k}_x^{i}+\mbox{k}_x^{s} ,\mbox{k}_y^{i}+\mbox{k}_y^{s}) \mbox{sinc}\left(L \Delta \mbox{k}_z/2\right)\mbox{exp}(i\Delta \mbox{k}_z L/2).
\label{eq:SPDCPUMPa}
\end{equation}
In this equation $L$ denotes the crystal length and $\Delta \mbox{k}_z = \mbox{k}_z^p-\mbox{k}_z^s - \mbox{k}_z^i $, with each $\mathbf{\mbox{k}}_z$ evaluated in terms of the vectors $\mathbf{k}_\bot$ using the adequate dispersion relation.

The product
\begin{equation}
\mathfrak{F}(\mbox{k}_x^s,\mbox{k}_y^s;\omega^s,\mbox{k}_x^i,\mbox{k}_y^i;\omega^i) =
\alpha(\omega^s + \omega^i)g F(\mbox{k}_x^s,\mbox{k}_y^s;\omega^s ,\mbox{k}_x^i,\mbox{k}_y^i;\omega^i), \quad\quad g =\mathcal{N}_p\mathcal{N}_s\mathcal{N}_i{\mathcal{X}},
\label{Eq:vecWJA}
\end{equation}
yields the probability amplitude to generate the idler and signal photons. It determines $\vert \Psi\rangle$ to first order in the perturbation theory. In general, $\mathfrak{F}(\mbox{k}_x^s,\mbox{k}_y^s;\omega^s,\mbox{k}_x^i,\mbox{k}_y^i;\omega^i)$
will be non negligible for continuous sets of the wave vectors $\mbox{k}_\bot$, so that the wave function cannot be  factorized, and entanglement in continuous and discrete polarization variables can be expected.

Equation (\ref{eq:SPDCPUMPa}) shows how the transverse momentum conservation condition ${\mathbf{k}}_\bot^p = {\mathbf{k}}_\bot^i +{\mathbf{k}}_\bot^s$ induces the transfer of the plane-wave spectrum from the pump beam to the two-photon state \cite{monken1}. If, for instance, one of the photons in the pair is projected in a state with a well defined
 value of $\mathbf{k}_\bot$, the other photon will have a  plane-wave spectrum proportional to the pump spectrum with an argument modified by a constant additive term. Note, however, that $\mathfrak{F}$ is also modulated both by  the longitudinal phase matching factor (which depends on ${\mathbf{k}}_\bot^i$ and ${\mathbf{k}}_\bot^s$ by the dispersion relations) and by the nonlinear response term ${\mathcal{X}}$. Thus,
the conditions under which the idler and/or signal photons inherit general features of the pump beam plane-wave spectrum are not evident.

An important function to calculate is the angular spectrum (AS), which describes the distribution of signal photons in the wave vector domain, and is defined as
\begin{equation}
\label{E:Rs}
R_s(\mathbf{k}_{0}^s)=\int d\omega^i \int d^2 k^i_\bot |\mathfrak{F}(\mbox{k}_{x,0}^s,\mbox{k}_{y,0}^s;\omega^s,\mbox{k}_x^i,\mbox{k}_y^i;\omega^i)|^2.
\end{equation}

The conditional angular spectrum (CAS), which is a function of $\mathbf{k}_\bot^{s}$ and $\mathbf{k}_{\bot,0}^{i}$, is defined as:
\begin{equation}
R_c (\mathbf{k}_{\bot}^{s},\mathbf{k}_{\bot,0}^{i};\omega^i_0)=\int d\omega^s  |\mathfrak{F}(\mbox{k}_{x}^s,\mbox{k}_{y}^s;\omega^s,\mbox{k}_{x,0}^i,\mbox{k}_{y,0}^i;\omega^i_0)|^2,
\label{E:Rc}
\end{equation}
and  represents the probability to detect an idler photon with wave vector $\mathbf{k}_{\bot,0}^{i}$ and frequency $\omega^i_0$ in coincidence with a signal photon with  wave vector $\mathbf{k}_\bot^{s}$.  Under a realistic  situation involving small but finite transverse dimensions of the pump beam and an usually wide but not so long crystal, there is a set of relevant pump wave vectors $\mathbf{k}^p$ that are close to satisfy the phase matching condition $\Delta \mbox{k}_z =0$ for a given idler wave vector $\mathbf{k}^i$.

\section{ SPDC for a propagation invariant pump with circular cylinder symmetry.}

The solutions of the scalar wave equation in free space that preserve its amplitude along a main propagation axis
are said to be propagation invariant. If the main propagation axis of the beam is chosen to be the $z$-axis, their scalar plane-wave spectrum $\tilde{\psi}_\kappa(\mbox{k}_x, \mbox{k}_y)$
is confined into a cone, that is, the transverse wave number  $\kappa_{\bot}^p$ of these beams, under ideal conditions, takes a unique value,
\begin{equation}
\tilde{\psi}_\kappa(\mbox{k}_x, \mbox{k}_y)= \Phi_\kappa(\varphi_{\mathbf{k}_\bot})\delta(\mbox{k}_{\bot} - \kappa_{\bot})/\kappa_\bot.
\end{equation}

Approximate realizations of propagation invariant beams in the laboratory correspond to superpositions of waves with vectors $\mathbf{k}$ in a narrow conic shape volume:
\begin{equation}
\delta(\mbox{k}_{\bot} - \kappa_{\bot})\rightarrow \frac{1}{W \sqrt{2 \pi}}\mbox{Exp}\left[- (\mbox{k}_{\bot} - \kappa_\bot)^2/2W^2\right],\quad \quad  \quad \quad \quad \quad W\ll \kappa_\bot.
\label{E:GB}
\end{equation}
with $W$ a waist parameter around the non-zero mean $\kappa_\bot$ value.

Propagation invariant beams with well defined orbital angular momentum are known as Bessel modes. The corresponding
angular spectra spectra is given by
\begin{equation}
\Phi_\ell(\varphi_{\mathbf{k}_\bot})=i^\ell \mbox{e}^{i \ell \varphi_{\mathbf{k}_\bot}},
\label{eq:besz}
\end{equation}
so that the modulus of $\tilde{\psi}_\kappa(\mbox{k}_x, \mbox{k}_y)$ is not dependent on $\varphi_{\mathbf{k}_\bot}$.
Their realizations in terms of narrow cones in wave vector space are called Bessel-Gauss modes.

Consider a  linearly polarized Bessel-Gauss mode generated in free space. The  beam is sent with its main direction of propagation along the $z$-axis, $i.$ $e.$, perpendicular to a birefringent crystal surface, and  with a polarization vector within the extraordinary plane. In the case the optical axis is taken as $\mathbf{a} = (0,\mathrm{a}_y,\mathrm{a}_z)$,  the angular spectra of the scalar pump beam ${\mathfrak{E}}^{sc}$ can be approximately expressed as \cite{footnote2}:
\begin{equation}
{\mathfrak{E}}^{sc}_{\mbox{k}_\bot^p,\ell,\omega}\sim
 {\cal E}^{sc}\frac{\mbox{e}^{-(\mbox{k}_{\bot}^p - \kappa_\bot^p)^2/2W^2 }}{\kappa_\bot^p}\mbox{e}^{i \ell \varphi_{\mathbf{k}_\bot^p}}{\hat{\mathbf{e}}_y}.\label{eq:scbessgauss}
\end{equation}
Inside the crystal the extraordinary mode that will give rise to the SPDC process evolves according to Eq.~(\ref{eq:bmodeE}) with an amplitude determined approximately by ${\mathfrak{E}}^{sc}\cdot {\mathfrak{E}}_\kappa$.
Notice that, in general the vectorial factors $\left[- (c^2/\epsilon_\bot\omega^2)\mathbf{k}^{\mbox{e}}({\bf a} \cdot \mathbf{k}^{\mbox{e} })+ {\bf a}\right]$ introduce anisotropy along the $z$ axis.

 For uniaxial media, the nonlinear optical susceptibilities $\chi_{r,l,q}^{(2)}$ are usually reported in the reference  frame where the birefringent axis is taken as the $z$-axis. In this frame we can evaluate ${\mathcal{X}}$, and then translate the results in terms of the rotated  wave vector ${\bf k}$. Let us take as a particular example the case of a beta barium borate (BBO) crystal and a type I phase-matching configuration. The symmetry of the crystal is such that, in the crystal natural frame ($\hat {\bf e}_3 = {\bf a}$):
\begin{equation}
{\mathcal{X}}\approx
-d_{22}\Big[ [\tilde {\bf e}_{x}^{\mbox{o}}\tilde{ \bf e}^{\mbox{o}}_{y} + \tilde {\bf e}_{y}^{\mbox{o}}\tilde {\bf e}^{\mbox{o}}_{x}]\tilde {\bf e}^{\mbox{e}}_{x}  +
 [\tilde{\bf e}_{x}^{\mbox{o}}\tilde{\bf e}^{\mbox{o}}_{x} - \tilde{\bf e}_{y}^{\mbox{o}}\tilde{\bf e}^{\mbox{o}}_{y}]\tilde{\bf e}^{\mbox{e}}_{y} \Big].
\end{equation}
 with $d_{22}\sim 2.2 \mathrm{pm/V}$  the element of the contracted  nonlinear matrix $d_{ij}$\cite{Dmitriev}.
In the case of vectorial electromagnetic beams in birefringent crystals, the components of the electric field
depend on the components of the wave vector, Eqs.~(\ref{eq:bmodeE}-\ref{eq:bmodeOO}), so that,
\begin{equation}
{\mathcal{X}}\approx -d_{22} \frac{\omega^s}{c}\frac{\omega^i}{c}\tilde{k}_z^p\Big[ [-\tilde{\mbox{k}}_y^s\tilde{\mbox{k}}_x^i-\tilde{\mbox{k}}_x^s\tilde{\mbox{k}}_y^i]\tilde{\mbox{k}}_x^p +
[\tilde{\mbox{k}}_y^s\tilde{\mbox{k}}_y^i-\tilde{\mbox{k}}_x^s\tilde{\mbox{k}}_x^i]\tilde{\mbox{k}}_y^p \Big],
\label{eq:chi}
\end{equation}
with $\tilde{\bf k}$ the rotated wave vector in the crystal reference frame.

For the system under consideration, if the crystal is wide and  the pump beam satisfies the paraxial condition, $\mbox{k}_\bot^p\ll\omega/c$, the conservation of transversal momentum makes reasonable that both the relevant ordinary and extraordinary modes are quasi parallel to the incident beam. All these considerations make feasible to replace the vectorial factor $\mathcal{X}$ by its effective value $\mathcal{X}_{eff}$. This implies that under these  conditions, the SPDC process of structured paraxial beams will be determined just by the scalar potential $\tilde\psi^p$ modulated by the longitudinal phase matching factor. Notice, however, that these considerations also suggest that the usage of vectorial non paraxial beams could open new perspectives in SPDC.

\subsection{SPDC angular spectrum and conditional angular spectrum for paraxial scalar Bessel beams.}

In this subsection we calculate  the AS, Eq.~(\ref{E:Rs}), and the  CAS, Eq.~(\ref{E:Rc}), functions for a quasi paraxial scalar Bessel pump beam, Eq.~(\ref{eq:scbessgauss}), under the conditions described in the last paragraph.

The first observation is that the modulus  of the  joint amplitude $\vert\mathfrak{F}\vert$, Eq.~(\ref{Eq:vecWJA}), in the scheme where Eq.~(\ref{eq:scbessgauss}) is valid, does not depend on $\ell$. So that the AS and CAS are $\ell$ independent \cite{fn}.

We assume  type-I SPDC in an uniaxial birefringent crystal  with its optical axis given by $\mathbf{a} = (0,\mathrm{a}_y,\mathrm{a}_z)$. For degenerated emission, i.e., $\omega^p= \omega=2\omega^s=2\omega^i$, the emitted photons dispersion relation is
\begin{equation}
\mbox{k}_{z,b}^{\mbox{o}}=\sqrt{\epsilon_\bot  \omega^2/4 c^2 - (\mbox{k}_{\bot}^b)^2}\quad\quad b=i,s,
\label{eq:dispord}
\end{equation}
with  the ordinary refraction index $n_o = \sqrt{\epsilon_\bot}$. The  dispersion relation for the pump wave which evolves in the extraordinary plane is
\begin{eqnarray}
\mbox{k}_z^{\tiny{E}}(\mathbf{k}_\bot,\omega)&=& -\beta \mathbf{a}\cdot\mathbf{k}_\bot  +\frac{\omega}{c}\mbox{n}_{eff}\sqrt{1
-\frac{\mbox{k}_\bot^2 c^2}{\omega^2}\eta},\label{eq:dispextraord}\\
\mbox{n}_{eff} &=&  \sqrt{\frac{\epsilon_\bot\epsilon_\parallel}{\epsilon_\bot + \Delta\epsilon{\rm a}_z^2}},\quad
\beta =\frac{\Delta\epsilon {\rm a}_z}{\epsilon_\bot + \Delta\epsilon{\rm a}_z^2},
\eta = \frac{1 }{\epsilon_\bot + \Delta\epsilon{\rm a}_z^2},
\end{eqnarray}
where $\Delta\epsilon= \epsilon_\parallel - \epsilon_\bot$.
 In the limit of normal incidence, this equation reduces to the  expression of the effective refractive index experienced by a paraxial pumping extraordinary wave,
$
\mbox{k}_z^{\tiny{E}}({\mathbf{k}}_\bot \sim \mathbf{0}) = \mbox{n}_{eff}\omega/c.
$
For quasi normal incidence, the deviation angle $\rho_{wo}$ of the Poynting vector with respect to the  wavefront inside the crystal can be approximated by $\tan\rho_{wo} \sim \beta{\rm a}_y$. So that $\beta$ measures the so called walk-off effect on the pump beam \cite{Walborn2004,torres,monken2008b,Walmsley2010}. The $\eta$ term in Eq.~(\ref{eq:dispextraord}) gives rise to  astigmatic effects \cite{monken2008}.

\subsubsection{SPDC angular spectrum.}

In order to obtain approximate  expressions for the AS function, we make a first order Taylor description of the phase mismatch term,
\begin{equation}
\Delta \mbox{k}_z\sim  \tilde{\kappa} - \mathbf{d} \cdot (\mathbf{k}_\bot^s +\mathbf{k}_\bot^i),
\label{E:MMTA}
\end{equation}
with
\begin{equation}
\tilde{\kappa} =(\omega/c) (n_{eff} -n_o) + (2 c/n_{o}\omega) (\mbox{k}_\bot^s)^2, \quad
\mathbf{d} = \beta \mathbf{a}_\bot + (2c/n_o \omega) \mathbf{k}_{\bot}^s.
\label{E:delta}
\end{equation}

 Writing the pump integration variable $\mathbf{k}_{\bot}^p$ in polar coordinates, and performing a rotation of the integration variable  by an angle $ \theta_r = {\rm arc} \cos(d_x/\vert\mathbf{d}\vert)$, the expression for the AS can be written in terms of a single integral:
\begin{equation}
R_s(\mbox{k}_{x}^s,\mbox{k}_{y}^s) \sim \vert g\alpha(\omega_p)\vert^2\mbox{e}^{
 -\sigma_{\rm{AS}}^{-2} \left(\left(\mbox{k}_{\bot}^s \right)^{2} -
r_{\rm{AS}}^2\right)^2  } \int_0^{2 \pi}  \mbox{Exp}\left[-\frac{(\gamma L)^2}{2} \left(\vert \mathbf{d} \vert \kappa_\bot^p \sin\varphi_p-\tilde{\kappa}\right)^2 \right]d\varphi_p,
\label{rs:integral}
\end{equation}
$$r_{\rm{AS}}^2 = \frac{1}{2}\left(\frac{ n_o\omega}{c} \right)^2\left(1-\frac{n_{eff}}{n_o}\right),\quad
\sigma_{\rm{AS}}^{-2}=  2(\gamma Lc/n_o\omega)^2. $$
Note that $r_{\rm{AS}}$ depends in general on the wavelength.
To obtain this expression, we have approximated the function $\mbox{sinc}(x)$ by a Gaussian function $\mbox{exp}[-(\gamma x)^2]$, $\gamma=0.4393$. We have also taken the limit $W\rightarrow 0$ with the restriction of a finite pump intensity.

According to Eq.~(\ref{rs:integral}), for $\kappa_\bot^p \ll (\omega/c) \vert n_{eff} -n_o\vert$, the AS is concentrated in a cone given by the condition  $\mbox{k}_{\bot}^s=(n_o\omega/\sqrt{2} c)\sqrt{ 1 -n_{eff}/n_o}=r_{\rm{AS}}$ for a negative birefringent crystal (a similar expression is obtained for Gaussian pump beams \cite{yasser2013}). The cone width is approximately given by $\sigma_{\rm{AS}}^{1/2}$.

Since
 \begin{equation}\tilde{\kappa} = \Big(\frac{2c}{n_o\omega}\Big)\Big( (k_\bot^{s})^2 -r_{\rm{AS}}^2\Big) \end{equation}
for $\kappa_\bot^p \ll (\omega/c) \vert n_{eff} -n_o\vert$, $\tilde{\kappa}\sim 0$.
That is, as expected, for  small values of $\kappa_\bot^p$ the angular spectrum will be similar to that obtained from Gaussian pump beams \cite{yasser2013}.

As $\kappa_\bot^p$ increases, the restriction $k_\bot^{s } \sim r_{\rm{AS}}$ is relaxed since the contribution
of the phase matching integral over the angle of wave vectors components of the pump beam in Eq.~(\ref{rs:integral})
\begin{equation} \int_0^{2 \pi}  \mbox{Exp}\left[-\frac{(\gamma L)^2}{2} \left(\vert \mathbf{d} \vert \kappa_\bot^p \sin\varphi_p-\tilde{\kappa}\right)^2 \right]d\varphi_p, \label{eq.int:as}\end{equation}
 is more relevant. This integral makes explicit that the phase matching condition involves the superposition effects of the wave vectors that arrive on the crystal in an isotropic way with respect to the crystal surface, but are not isotropically distributed with respect to the optical axis. In fact, this integral
gives rise to anisotropic effects in the AS due to the dependence of
 \begin{equation}
 \vert \mathbf{d} \vert = \Big(\frac{2c}{n_o\omega}\Big)\Big[ k_x^{s2} + \Big(k_y^s + \frac{ n_o\omega \beta{\rm a}_y}{2c}\Big)^2\Big]^{1/2}\end{equation}
  on the orientation of the axis $\mathbf{a}=(0,\rm{a}_y,\rm{a}_z)$.  This dependence suggests that
 the effects  of this integral on the AS  will yield structures that are not centered at the origin, but displaced in the direction of
 the $Y$-axis. This displacement would be absent if $\kappa_\bot^p$ were zero, $e.$ $g.$, for a Gaussian pump beam. Notice that $\kappa_\bot^p$ appears in the integrand in a combination $\kappa_\bot^p \sin\varphi_p$. So that, depending on the sign of the sine function in the exponent, the  values of $\tilde \kappa$ over which the  integral is relevant will be either positive (yielding greater values of $k_\bot^s$ with respect to $r_{\rm{AS}}$) or negative (yielding lower values of $k_\bot^s$ with respect to $r_{\rm{AS}}$). From these considerations we expect that the structure of the AS will now involve two non homogenous and non concentric cones with different radii. The actual displacement of the center of these cones can be estimated by evaluating the zeros of the exponent in the integral given in Eq.~(\ref{eq.int:as}) for the particular case $k_x^s=0$. We obtain that, for a negative birefringent crystal and for $ \vert n_o\omega \beta{\rm a}_y/2c\vert\sim r_{\rm{AS}} \gg \kappa_\bot^p$, the  cone with quasi circular transverse structure and the biggest   radius $R_+$ has an axis that passes through the point $(0,A_+)$ with
  \begin{eqnarray}
 R_+ \sim r_{\rm{AS}} &-& \frac{\kappa_\bot}{2}\Big(1 + \frac{ n_o\omega \beta{\rm a}_y}{2cr_{\rm{AS}}} - \frac{\kappa_\bot}{2r_{\rm{AS}}}\Big),\nonumber\\
 A_+ \sim &-& \frac{\kappa_\bot}{2}\Big(1 + \frac{ n_o\omega \beta{\rm a}_y}{2cr_{\rm{AS}}} - \frac{\kappa_\bot}{2r_{\rm{AS}}}\Big), \label{eq:r+}
 \end{eqnarray}
 while the axis of the cone with lowest transverse radius $R_-$  passes through  $(0,A_-)$ with
   \begin{eqnarray}
 R_- \sim r_{\rm{AS}} &-& \frac{\kappa_\bot}{2}\Big(1 - \frac{ n_o\omega \beta{\rm a}_y}{2cr_{\rm{AS}}} + \frac{\kappa_\bot}{2r_{\rm{AS}}}\Big),\nonumber\\
 A_-         \sim &+&\frac{\kappa_\bot}{2}\Big(1 + \frac{ n_o\omega \beta{\rm a}_y}{2cr_{\rm{AS}}} - \frac{\kappa_\bot}{2r_{\rm{AS}}}\Big). \label{eq:r-}
 \end{eqnarray}
 The two emission cones almost touch each other along the direction defined by the wave vector $\sim (0,-r_{\rm{AS}} +\kappa/2,k_z)$ for an optical axis $\mathbf{a}=(0,\rm{a}_y,\rm{a}_z)$. If the optical axis were located at the $X$--$Z$ plane the direction at which the cones would touch each other
 would be defined by a wave vector contained in the $K_x$- $K_z$ plane.  The double conical structure of the AS reflects both  the anisotropy  of the wave vectors in the incoming Bessel pump beam with respect to the optical axis  and walk-off effects encoded in $ \beta{\rm a}_y$ term of the extraordinary beam dispersion relation.

\begin{figure*}
\begin{center}
\subfloat[]{\label{F:ASL1}\includegraphics[width=0.6\textwidth]{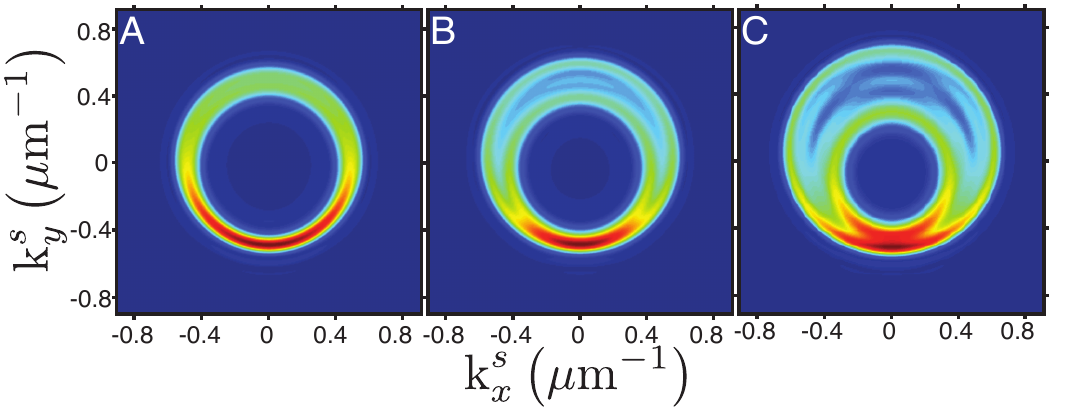}}\newline
\subfloat[]{\label{F:ASL1Expr}\includegraphics[width=0.6\textwidth]{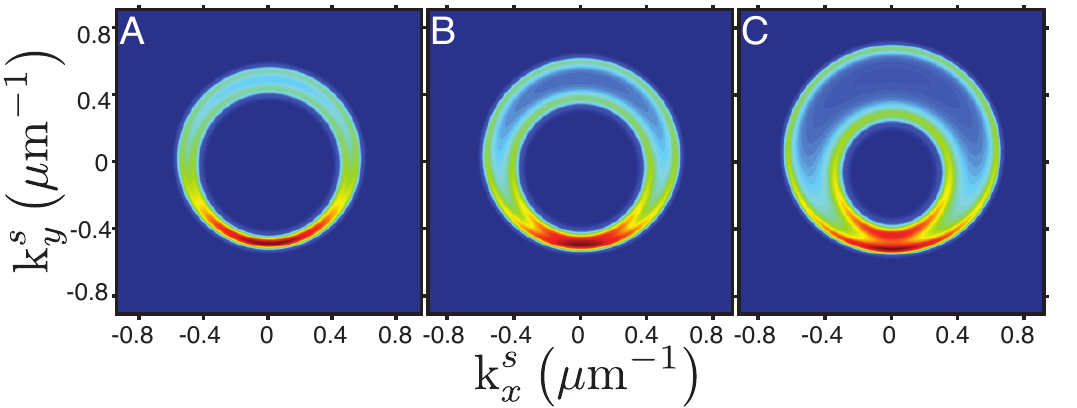}}\newline
\subfloat[]{\label{F:ASL2}\includegraphics[width=0.6\textwidth]{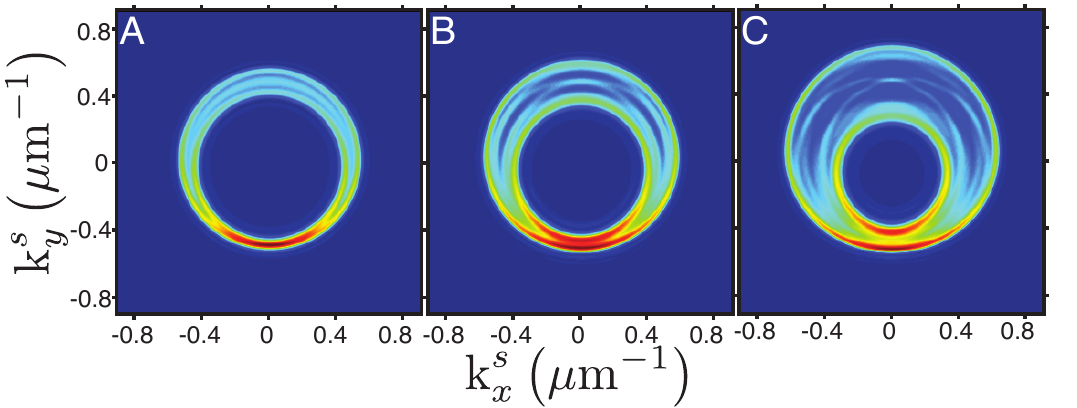}}\newline
\subfloat[]{\label{F:ASL2Expr}\includegraphics[width=0.6\textwidth]{ASBsslExprL1.pdf}}\newline
\end{center}
\caption{Angular spectrum (AS) for  a pump beam with a waist $W=0.0007 \mu$m$^{-1}$ and different values of transverse wave number: (A) $\kappa_\bot^p=0.05 \mu$m$^{-1}$; (B)  $\kappa_\bot^p=0.09 \mu$m$^{-1}$; (C) $\kappa_\bot^p=0.15 \mu$m$^{-1}$. Figures (a) and (b) consider a $1$ mm long BBO crystal while in (c) and (d) the crystal length is increased to 2mm. In figures (a) and (c), the AS is calculated numerically through Eq.(\ref{Eq:vecWJA}), while in figures (b) and (d) the AS is calculated from the analytic expression Eq.(\ref{rs:integral}). The optical axis of the crystal is located in the $Y$-$Z$ plane  $\mathbf{a} \sim (0,0.49,0.87)$. For normal incidence the resulting walk-off angle is  $\rho_{wo} \sim -\beta{\rm a}_y\sim 0.068 \rm{rad}$ while the emission
cones have a aperture angles around $theta_{A} \sim r_{\rm{AS}}/k^s_z \sim 0.034\rm{rad}$, $r_{\rm{AS}} =\sim 0.49\mu$m$^{-1}$. The estimated transverse radii of the major emission cones are $R_+^A\sim 0.051\mu$m$^{-1}$, $R_+^B\sim 0.053\mu$m$^{-1}$ and $R_+^C\sim 0.056\mu$m$^{-1}$, while for the minor emission cones are $R_-^A\sim 0.042\mu$m$^{-1}$,$R_-^B\sim 0.036\mu$m$^{-1}$ and $R_-^C\sim 0.027\mu$m$^{-1}$; their centers are located at $A_\pm^A\sim \pm0.02\mu$m$^{-1}$, $A_\pm^B\sim\pm0.04\mu$m$^{-1}$ and $A_\pm^C\sim\pm 0.07\mu$m$^{-1}$. }
\end{figure*}

Our theoretical scheme has been implemented using parameters from the reported experimental setup in Ref.~\cite{uren2012}. We consider a SPDC source based on a BBO crystal, cut for type-I phase matching for degenerated emission and normal incidence of a pump quasi plane wave with wavelength centered at $406.8$ nm ($\theta_\mathbf{a}=29.3^\circ$, optical axis $\mathbf{a} = (0,\sin\theta_{\bf a},\cos\theta_{\bf a})$). The numerical simulations consider monochromatic Bessel-Gauss pump beams with a waist $W=0.0007 \mu$m$^{-1}$ and three different values for the transverse wave number parameter $\kappa_\bot^p=0.05,0.09,0.15\mu$m$^{-1}$. They are performed both using the complete expression of the transition rate Eq.(\ref{E:Rs}) without resorting approximations to the dispersion relations, Figs.~\ref{F:ASL1}-\ref{F:ASL2}, and using the analytic expression, Eq.~(\ref{rs:integral}), Figs.~\ref{F:ASL1Expr}-\ref{F:ASL2Expr}, for two different crystal lengths. The AS function for $\kappa_\bot^p=0.05\mu$m$^{-1}$ shows an asymmetry compatible with the experimental results reported in Ref.\cite{uren2012}. In this case, the AS function is concentrated in a cone of radius $0.49\mu$m$^{-1}$ and width between $\sim 0.01 -0.03\mu$m$^{-1}$ dependent on $\mbox{k}_{y}^s/\mbox{k}_{x}^s$. Both radius and width are within the expectations described  above. As $\kappa_\bot^p$ increases, the anisotropy is more visible and the spectrum is better described by two non concentric cones that almost touch each other along a line determined by the orientation of the optical axis of the nonlinear crystal. This is illustrated in Fig.(\ref{F:ASL1})B where $\kappa_\bot^p=0.09\mu$m$^{-1}$ and Fig.(\ref{F:ASL1})C with $\kappa_\bot^p=0.15\mu$m$^{-1}$.
As the crystal length is increased the regions where the AS has significant values are smaller and the anisotropy associated to the extraordinary pump beam dispersion relation is more evident. This, along with the oscillatory behavior of the sinc function, leads to a higher structured landscape for the AS function, which, nevertheless is still concentrated in two narrow non concentric cones. The radius of the external cone increases as $\kappa_\bot$ increases while the radius of the internal cone decreases as $\kappa_\bot^p$ decreases; all this is in accordance to the expectations described above. By comparing Figs.~\ref{F:ASL1} with Figs.~\ref{F:ASL1Expr} and Figs.~\ref{F:ASL2} with Figs.~\ref{F:ASL2Expr} we can conclude that the analytic expression, Eq.~(\ref{rs:integral}), reproduces the general features of the AS function for $\kappa_\bot^p$ at least as high as 0.15$\mu$m$^{-1}$. The displacement of the
centers of the AS cones and their radii are also correctly estimated by Eqs.~(\ref{eq:r+}-\ref{eq:r-}).

\subsubsection{SPDC conditional angular  spectrum.}
Now we shall study the probability to detect an idler photon with wave vector $\mathbf{k}_{\bot,0}^{i}$ in coincidence with a signal photon with  wave vector $\mathbf{k}_\bot^{s}$ in terms of the conditional angular  spectrum,  Eq.~(\ref{E:Rc}).
Using the same approximations for the phase matching function and phase mismatch term, it is possible to obtain an useful expression for the CAS function. In this case we do not take the limit $W\rightarrow 0$ to make more explicit the role of this parameter in the expected CAS to be measured in the laboratory. Replacing Eq.~(\ref{eq:SPDCPUMPa}),  Eq.~(\ref{eq:scbessgauss}) and Eq.~(\ref{E:MMTA}) in Eq.~(\ref{E:Rc}), for degenerate SPDC, we obtain:
\begin{equation}
R_c (\mathbf{k}_{\bot}^s,\omega^s;\mathbf{k}_\bot^i, \omega^i)\sim  \frac{\vert g\alpha(\omega^p)\vert^2}{2\pi W^2}
\mbox{e}^{-(\vert\mathbf{k}_\bot^s -\mathbf{K}^0_\bot\vert^2 -\mathcal{R}_{\mathbf{k}}^2)/2W_{eff}^2},
\label{E:F2}
\end{equation}
with
\begin{eqnarray}
W_{eff}^{-2} &=& W^{-2} + \gamma^2 L^2\beta^2(\mathbf{a}\cdot\hat{\mathbf{k}}_\bot)^2,\nonumber\\
\mathbf{K}^0_\bot &=& -\frac{W_{eff}^{2}}{W^{2}}\mathbf{k}_\bot^i + (\gamma L W_{eff})^{2}\Big(\frac{\omega}{c} (n_{eff} -n_o) - \beta\mathbf{a}\cdot\mathbf{k}_{\bot}^i\Big)\Big(\frac{2c}{n_o\omega}
\mathbf{k}_{\bot}^i -\beta\mathbf{a}\Big),\nonumber\\
\mathcal{R}_{\mathbf{k}}^2&=& \frac{W_{eff}^{2}}{W^{2}}(\kappa_\bot^p)^2 + (\mbox{k}_{\bot}^i)^2\Big[
\frac{W_{eff}^{4}}{W^{4}} -\frac{W_{eff}^{2}}{W^{2}} \Big] + 2(\gamma LW_{eff})^2\Big(\frac{\omega}{c} (n_{eff} -n_o) - \beta\mathbf{a}\cdot\mathbf{k}_{\bot}^i\Big)\Big(\frac{2c}{n_o\omega}
(\mbox{k}_{\bot}^i)^2 -\beta\mathbf{a}\cdot\mathbf{k}_\bot^i\Big) \nonumber\\ &-& \Big(\frac{2c}{n_o\omega}
(\mbox{k}_{\bot}^i)^2 -\beta\mathbf{a}\cdot\mathbf{k}_{\bot}^i\Big)^2(\gamma L W_{eff})^{2}\Big[1 - (\gamma L W_{eff})^{2} \vert \frac{2c}{n_o\omega}\mathbf{k}_{\bot}^i -\beta\mathbf{a}_\bot\vert^2  \Big].
\end{eqnarray}

If the crystal is not too long $W_{eff} \sim W$, $\mathbf{K}^0_\bot= - \mathbf{k}_{\bot}^i$
and $\mathcal{R}_{\mathbf{k}}\sim k^{0}_\bot$; as the length of the crystal increases the longitudinal phase matching condition becomes more relevant and two types of corrections arises. One of them is related to the
differences between the extraordinary and ordinary refractive indices and the other includes effects of the orientation of the birefringent axis $\mathbf{a}$. The latter is  highly anisotropic and is directly related to the spatial walk-off. Both effects make evident that the idler photon characteristics can induce observable differences between the general characteristics of the signal photon with respect to the structured pump beam.

 Since the condition of
propagation invariance of a mode corresponds to restrict its wave vectors into a non necessarily homogenous cone, we observe that, whenever $\mathcal{R}_{\mathbf{k}}^2>0$,  the structure of Eq.~(\ref{E:F2}) is  that of an approximate propagation invariant signal photon, as reported in reference \cite{uren2012}.
In the idealized limit of $W\rightarrow 0$,
$$R_c (\mathbf{k}_{\bot}^s,\omega^s;\mathbf{k}_{\bot}^i, \omega^i)\rightarrow  \frac{\vert g\alpha(\omega_p)\vert^2}{k_\bot^0 }
\delta(\vert\mathbf{k}_{\bot}^s+\mathbf{k}_{\bot}^i\vert - \kappa_\bot^p)\times $$
\begin{equation}
\mbox{Exp}\left[-\gamma^2 L^2\left(2(\mathbf{k}_{\bot}^i+\mathbf{k}_{\bot}^s)\cdot
(\frac{2c}{n_o\omega} \mathbf{k}_{\bot}^i -\beta\mathbf{a})
(\frac{\omega}{c} (n_{eff} -n_o) - \beta\mathbf{a}\cdot\mathbf{k}_{\bot}^i)
+\frac{\omega}{c} (n_{eff} -n_o) - \beta\mathbf{a}\cdot\mathbf{k}_{\bot}^i\right)^2\right].
\end{equation}
In this limit, the condition of propagation invariance for a beam with main propagation axis along $(2c/\omega n_o)(-\mathbf{k}_{\bot}^i,\mbox{k}_{z}^i)$ can be written as
\begin{equation}
\Big[\frac{\mbox{k}_{z}^i}{\mbox{k}_{\bot}^i \mbox{k}_{0}^i}(\mathbf{k}_{\bot}^i\cdot \mathbf{k}_{\bot}^s)+\frac{\mbox{k}_{\bot}^i\mbox{k}_{z}^s}{\mbox{k}_{0}^i}\Big]^ 2 +\Big[\frac{\mathbf{{k}}_{\bot}^i\times \mathbf{{k}}_{\bot}^s}{\mbox{k}_{\bot}^i} \Big]^2
= (\kappa_\bot^p)^2, \quad\quad \mbox{k}_{0}^i =n_o\omega^i/c=\vert\mathbf{k}^i\vert;
\end{equation}
expression that can also be written in the form
\begin{equation}
\vert\mathbf{k}_{\bot}^s+\mathbf{k}_{\bot}^i\vert^2\Big(\frac{\omega^i}{\omega^s}\Big)^2
+(\mbox{k}_\bot^s)^2\Big[  1 - \Big(\frac{\omega^i}{\omega^s}\Big)^2\Big] -\frac{(\mbox{k}_{\bot}^s)^2}{(\mbox{k}_0^i)^2}\Big[
(\mathbf{k}_{\bot}^i\cdot\hat{\mathbf{k}}_{\bot}^s)^2 + (\mbox{k}_\bot^i)^2 +\mathbf{k}_\bot^i\cdot\mathbf{k}_\bot^s\Big]
= \kappa_\bot^2.\end{equation}
As a consequence, the degenerate SPDC leads to approximate propagation invariant photons whenever $\mbox{k}_\bot^s\ll \vert\mathbf{k}^s\vert$. Under the experimental  conditions reported in Ref.~\cite{uren2012} $\mbox{k}_\bot^s/\vert\mathbf{k}^s\vert\sim 0.1$.

As we have shown, for paraxial pump beams, the mean radii of the AS cone is determined by the difference
between refractive indices, while the radius $\kappa_\bot$ of the modes  that describe the  emitted photons coincide with that of the pump Bessel beam.

\begin{figure*}
\begin{center}
\subfloat[]{\label{F:CASL1}\includegraphics[width=0.6\textwidth]{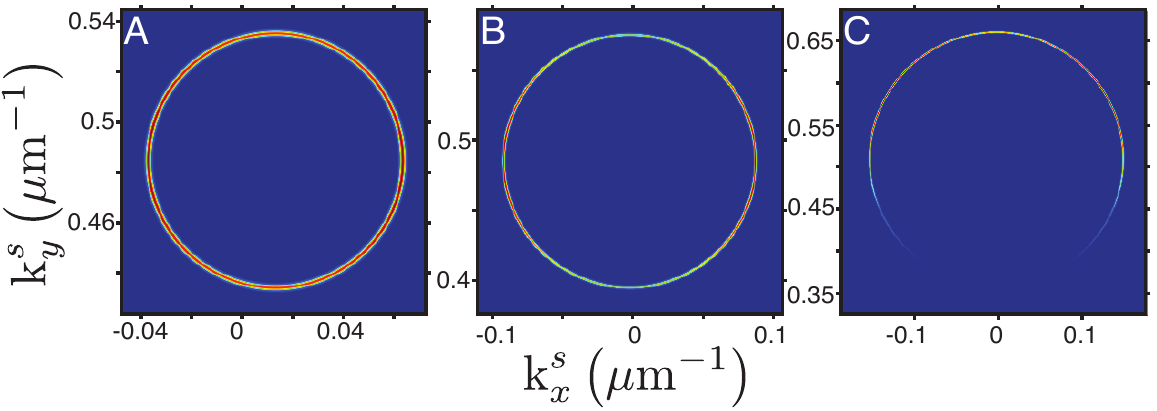}}\newline
\subfloat[]{\label{F:CASL2}\includegraphics[width=0.6\textwidth]{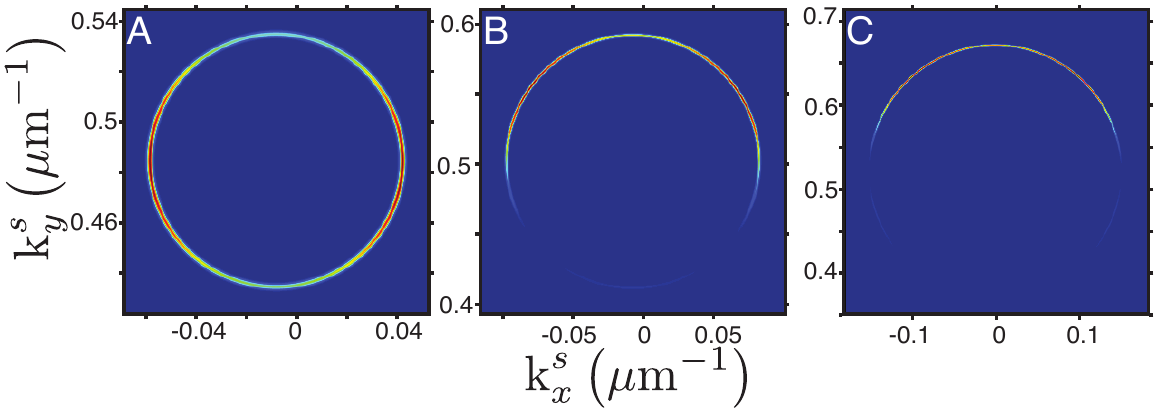}}\newline
\subfloat[]{\label{F:CAS100}\includegraphics[width=0.6\textwidth]{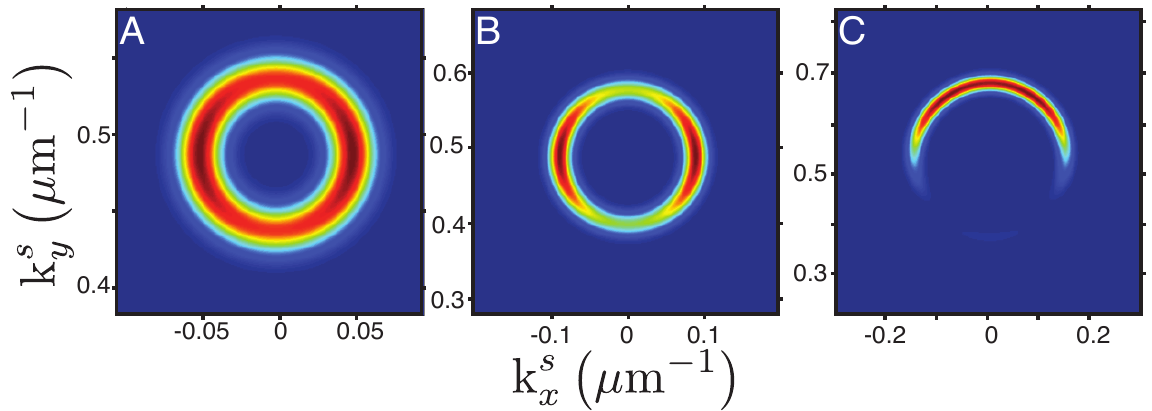}}\newline
\end{center}
\caption{Maximal conditional angular spectrum for a pump Bessel Gauss beam with waist $W=0.0007 \mu$m$^{-1}$  in Figs.~(a) and (b) and $W=0.007 \mu$m$^{-1}$ in Fig.~(c).The transverse wave numbers are: (A)  $\kappa_\bot^p=0.05 \mu$m$^{-1}$, (B)  $\kappa_\bot^p=0.09 \mu$m$^{-1}$, and (C)   $\kappa_\bot^p=0.15 \mu$m$^{-1}$. The crystal length is $1
$ mm in Figs.~(a) and (c) and 2 mm in Fig.~(b). The optical axis of the crystal is located in the $Y$-$Z$ plane  $\mathbf{a} \sim (0,0.49,0.87).$}
\end{figure*}

In Fig.~(2),  the CAS function is illustrated using the same general parameters as in Fig.~(1). We take the transversal wave vector of the idler photon $\mathbf{k}_{\bot}^{i,max}$ that maximizes the counts in the AS, and name it "maximal  CAS".
The results found for the AS let us expect that the maximum probability of emission of an idler photon would be obtained for $\mathbf{k}_{\bot,0}^{i,max}\sim (0,-r_{\rm{AS}})$ where the two AS cones almost touch each other. In the case reported in  Fig.~(\ref{F:ASL1})A, the maximum AS corresponds to $\mathbf{k}_{\bot}^{i,max}=(0.027,-0.485)\mu$m$^{-1}$ and Fig.~(\ref{F:CASL1}) shows the  corresponding maximal CAS. The transverse phase matching condition guarantees that the maximum probability signal photon wave vector will be located nearby the wave vector $-\mathbf{k}_{\bot,0}^{i,max}$ within a radius $\sim \kappa_\bot$. This region of the AS is where the two cones have the greatest separation. The propagation invariance structure of the pump beam makes that the CAS will also have an annular  shape for $\mathbf{k}_{\bot}^{s,max}$ with a radius $\sim\kappa_\bot$. Since the AS nearby $-\mathbf{k}_{\bot,0}^{i,max}$ is quite asymmetric, the maximal CAS is also expected to be an asymmetric ring. In fact as $\kappa_\bot$ increases, it may happen that the width and separation of the AS cones nearby the signal photon location are smaller than $\kappa_\bot$. Then the signal photon ring will not even close.
 The width of the ring is approximately equal to the width in wave vector space of the incident beam with an slight dependence on the crystal length, see Fig.~(\ref{F:CASL2}).  Fig.~(\ref{F:CAS100}) illustrates the maximal CAS for an incident pump beam with a waist $W=0.007\mu$m$^{-1}$; it shows in greater detail its anisotropic transversal structure as a function of $\kappa_\bot^p$. The similarity with the experimental results reported in Ref.\cite{uren2012} is also evident.

Summarizing, the intensity structure of the maximal conditional spectrum is concentrated in a cone with an anisotropy that increases as $\kappa_\bot^p$ increases. As a result, most conditionally emitted signal photons will be approximately propagation invariant but they will not correspond, in general, to Bessel-Gauss photons with a well defined $\ell$ value. Signal photons might be better approximated by Bessel-Gauss photons when paraxial pump beams are used.

\subsection{Post selected photons with well defined angular momentum along different propagation axes.}
Taking into account the results obtained in last section, it is relevant to study the angular momentum correlations of
signal and idler photons. Thus, in this subsection, we  consider the same general set up described in the previous subsection, and now evaluate the emission probability of photon pairs each of which has a Bessel-Gauss  structure. The calculation is made by a direct integration of Eq.~(\ref{E:St}) with the structured pump beam described by Eq.~(\ref{eq:bmodeE}), and the idler and signal photons with the proper structure factor
\begin{equation}
\label{eq:bmodeO}
{\mathfrak{E}}_{{\mbox{o}},\kappa}(\mbox{k}_x,\mbox{k}_y;\omega)= {\bf a} \times {\bf k}^{\mbox{o}}\tilde\psi_\kappa(\mbox{k}_x,\mbox{k}_y).
\end{equation}

\begin{figure}
\begin{center}
{\includegraphics[width=0.6\textwidth]{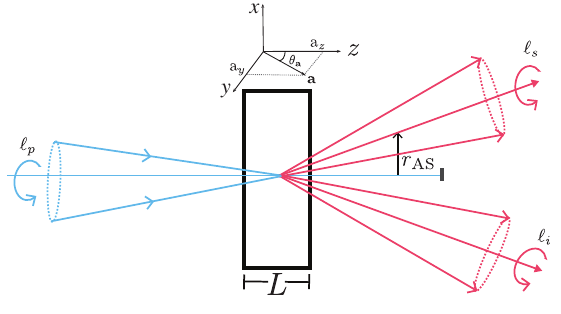}}
\caption{Schematic picture of the SPDC process involving  propagation invariant pump, signal and idler photons with orbital angular momentum.  The emission cone is expected to have a radii given by $r_{\rm{AS}}=\left( n_o\omega/c \right)^2\left(1-n_{eff}/n_o\right)/2$. The main propagation axis of the signal and idler photons will be approximately located on that cone.}\label{F:diag}
\end{center}
\end{figure}

According to the results of last section, the main propagation axis of the post selected   signal  and  idler Bessel photons is expected to differ from the pump beam axis. For a pump field  close to satisfy the paraxial approximation, $\kappa_\bot^p\ll k_z^p$, this joint probability is expected to be maximal for photons with their main propagation axis nearby the cone with squared radius $r_{\rm{AS}}^2 =\left( n_o\omega/c \right)^2\left(1-n_{eff}/n_o\right)/2$ in wave vector space. In Fig.~(\ref{F:diag}), a schematic picture of the  SPDC process under study in this section is shown.

  Given a scalar Bessel beam with main propagation axis
\begin{equation}
\hat{p}_3=(\sin\tilde\theta\cos\tilde\varphi,
\sin\tilde\theta\sin\tilde\varphi,\cos\tilde\theta),\label{eq:p3}
\end{equation}
the  vectors
\begin{eqnarray}
\hat{p}_1&=&(\cos\tilde\theta\cos\tilde\varphi,
\cos\tilde\theta\sin\tilde\varphi,-\sin\tilde\theta),\\
\hat{p}_2&=&(-\sin\tilde\varphi,\cos\tilde\varphi,0),
\end{eqnarray}
together with (\ref{eq:p3}) form an orthogonal basis on which any vector ${\bf{k}}$, with components $(\mbox{k}_x,k_y,k_z)$ in the frame where the normal of the surface of the crystal coincides with the $z$-axis, has components
\begin{eqnarray}
\tilde{\mbox{k}}_x &=& \mbox{k}_x\cos\tilde\theta\cos\tilde\varphi + \mbox{k}_y\cos\tilde\theta\sin\tilde\varphi - \mbox{k}_z\sin\tilde\theta\nonumber\\
\tilde{\mbox{k}}_y &=&-\mbox{k}_x\sin\tilde\varphi + \mbox{k}_y\cos\tilde\varphi\\
\tilde{\mbox{k}}_z &=&\mbox{k}_x\sin\tilde\theta\cos\tilde\varphi+\mbox{k}_y\sin\tilde\theta\sin\tilde\varphi + \mbox{k}_z\cos\tilde\theta.
\end{eqnarray}

 In the basis $\{\hat{p}_1,\hat{p}_2,\hat{p}_3\}$ the scalar factor of the Bessel mode is
\begin{equation}
\psi_{\hat{p_3};\kappa_\bot,\ell}(\tilde{\bf r} ) =\int d^3\tilde{k} e^{i\tilde{\bf{k}}\cdot\tilde{\bf r}}\delta(\tilde{k}_z - \tilde{K}_z(\omega,\tilde{\vec{k}}_\bot))
\frac{\delta(\tilde{k}_\bot -\kappa_\bot)}{\kappa_\bot}i^\ell e^{i\ell\tilde\varphi}
\end{equation}
with $\tilde{K}_z$ given by the adequate dispersion relation, Eqs.~(\ref{eq:dispord}- \ref{eq:dispextraord}).
Writing the vector $\tilde{\bf k}$ in terms of $(\mbox{k}_x,\mbox{k}_y,\mbox{k}_z)$, and using that $\tilde{\bf{k}}\cdot\tilde{\bf r}=\bf{k}\cdot\bf{r}$, we can get an expression for
$\tilde \psi_{\hat{p}_3;\kappa_\bot,\ell}(\mbox{k}_x,\mbox{k}_y)$. For an ordinary Bessel mode it results

\begin{eqnarray}
\tilde \psi_{\hat{p_3};\kappa_\bot,\ell}^{(O)}({\bf k} ) &=& \pm\frac{k_\bot^{\vert\ell\vert}}{\kappa_\bot^{\vert\ell\vert}}\frac{\delta({k}_z - K_z^{(O)}(\omega_B,\kappa_\bot,\tilde\theta;\varphi) )}{\kappa_\bot}
  \frac{\delta(k_\bot - K_\bot^{(O)}(\kappa_\bot,\omega_B,\tilde\theta;\varphi))}{{\rm Jac}^{(O)}(\kappa_\bot,\omega_B,\tilde\theta;\varphi))}\nonumber\\
&\times&\sum_{m=0}^{\vert\ell\vert}\Big( \begin{matrix} \vert\ell\vert\\ \vert\ell\vert -m \end{matrix}\Big) \Big(\cos\tilde\theta\cos(\varphi-\tilde\varphi)+ i\frac{\ell}{\vert\ell\vert} \sin(\varphi-\tilde\varphi)\Big)^m\Big(-\frac{k_z}{k_\bot}\sin\tilde\theta\Big)^{\vert\ell\vert -m}\label{eq:besoa}.
\end{eqnarray}
$$K_z^{(O)}(\kappa_\bot,\omega_B;\varphi;\tilde\theta;\tilde\varphi) =\Big[\frac{\sqrt{\epsilon_\bot\omega_B^2/c^2 -\kappa_\bot^2}}{\cos\tilde\theta}$$
\begin{equation}
\pm \vert \tan\tilde\theta\cos(\tilde{\varphi} -\varphi)\vert\sqrt{\kappa_\bot^2/\cos^2\tilde\theta -(\epsilon_\bot\omega_B^2/c^2)\tan^2\tilde\theta\sin^2(\tilde{\varphi} -\varphi)}\Big]/
(1 +\tan^2\tilde\theta\cos^2(\tilde{\varphi} -\varphi)),\label{eq:effdisrel}
\end{equation}

\begin{equation}
K_\bot^{(O)}(\kappa_\bot,\omega_B;\varphi;\tilde\theta;\tilde\varphi)=\sqrt{\epsilon_\bot \omega_B^2/c^2 - K_z^{(O)2}},\label{eq:sis2}
\end{equation}
and the Jacobian term is
\begin{equation}
\rm{Jac}^{(O)}(\omega_B;k_\bot;\varphi;\tilde\theta,\tilde\varphi))=\frac{\vert k_\bot-k_z\cos(\tilde\varphi -\varphi)\sin\tilde\theta\vert}{\kappa_\bot}. \label{eq:Jac}
\end{equation}

The dependence of the relevant values of $\mbox{k}_z$ and $\mbox{k}_\bot$, $K_z^{(O)}$ and $K_\bot^{(O)}$, on the angle $\varphi$ reflects the elliptic shape of the beam in the  wave vector plane defined by a constant value of  $\mbox{k}_z$.
 The ordinary dispersion relation term $\sqrt{\epsilon_\bot\omega_B^2/c^2 -\kappa_\bot^2}$ is a direct measure of the electromagnetic momentum of the Bessel beam along the $\hat p_3$ axis. $K_z^{(O)}$ is proportional to the momenta along the $z$-axis; $\kappa_\bot$ determines directly the radial momenta  of ordinary photons perpendicular to the $\hat p_3$ axis; the Jacobian terms Eq.~(\ref{eq:Jac}) are a consequence of the conceptual difference between $k_\bot$ and $\kappa_\bot$. Note that the effective dispersion relation for the Bessel mode, Eq.~(\ref{eq:effdisrel}) gives real values just for values of the angular variable $\varphi$ satisfying
 $\vert\kappa_\bot\vert\ge\vert(n_o\omega_B/c)\sin\tilde\theta\vert$.
 The summation terms in Eq.~(\ref{eq:besoa}) substitute the $e^{i\ell\varphi}$ term that would arise for a Bessel
beam propagating along the $\hat e_3$-axis. Notice that similar terms  arise in the  description of SPDC for other beams exhibiting orbital angular momentum as, for instance, Laguerre Gaussian beams \cite{Osorio2008}.

The transition amplitude of the generation of a signal Bessel photon with angular momenta $\ell^s$ and transverse wave number $\kappa_\bot^s$ along with  an idler Bessel photon with angular momenta $\ell^i$ and transverse wave number $\kappa_\bot^i$ is proportional to
$$
F(\ell^p, \kappa_\bot^p,\omega^p; \ell^s, \kappa_\bot^s,\omega^s;\ell^i, \kappa_\bot^i,\omega^i ) =$$
\begin{equation}
\int d^3\mathbf{k}^s\int d^3\mathbf{k}^i\tilde \psi_{\hat{e_3};\kappa_\bot^p,\ell^p}^{(E)}(\mathbf{k}_\bot^i +\mathbf{k}_\bot^s;\ell^p)\tilde \psi_{\hat{p}^s;\kappa_\bot^s,\ell^s}^{(\mbox{o})}({\bf k}_\bot^s )\tilde \psi_{\hat{p}^i;\kappa_\bot^i,\ell^i}^{(\mbox{o})}(\mathbf{k}_\bot^i) \rm{sinc}(L\Delta k_z/2);
\label{eq:Flslilp}
\end{equation}
the proportionality factor $g =\vert \alpha\vert\mathcal{N}_p\mathcal{N}_s\mathcal{N}_i{\mathcal{X}_{eff}}$ involves the pump coherent amplitude $\alpha$, the normalization factors of the pump, idler and signal photons as well as the adequate effective nonlinear susceptibility.

According to the results of last section,  the transition amplitude will become maximal if the signal photon has an orientation axis $\hat p_3^{(s)}$ determined by the maxima in the AS function, while the orientation of the idler photon corresponds to $\tilde\theta_i=\tilde\theta_s$ and $\tilde\varphi_i = \tilde\varphi_s + \pi$ provided that $\kappa_\bot^s<<\kappa_\bot^i$. The anisotropy of the AS has as a consequence that the general features of the
photon pairs depend also on their emission orientation. 
Notice also that in this scheme the idler and signal photons could be distinguished by their perpendicular wave vector.

\begin{figure*}
\begin{center}
\subfloat[]{\label{F:L0-MPI-01}\includegraphics[width=0.32\textwidth]{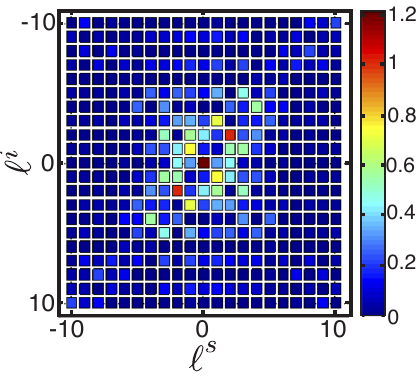}}
\subfloat[]{\label{F:L1-MPI-01}\includegraphics[width=0.32\textwidth]{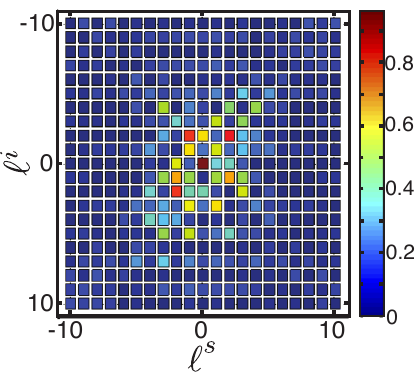}}
\subfloat[]{\label{F:L2-MPI-01}\includegraphics[width=0.32\textwidth]{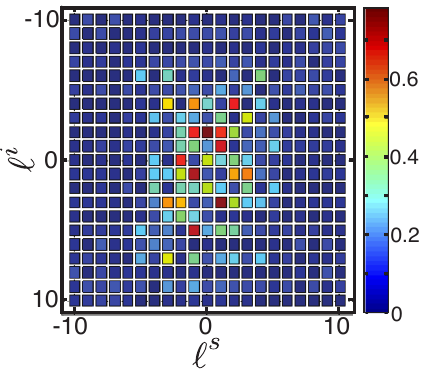}}\newline
\subfloat[]{\label{F:L0-0PI-01}\includegraphics[width=0.32\textwidth]{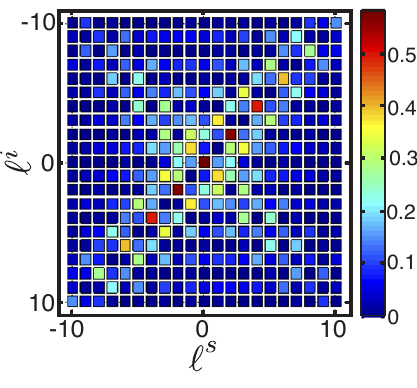}}
\subfloat[]{\label{F:L1-0PI-01}\includegraphics[width=0.32\textwidth]{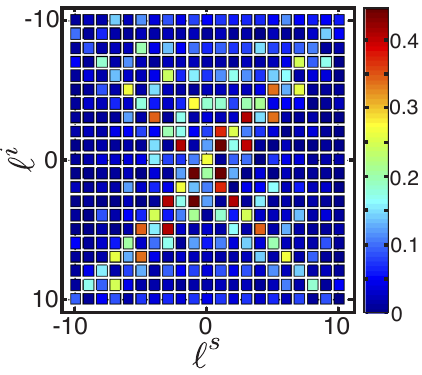}}
\subfloat[]{\label{F:L2-0PI-01}\includegraphics[width=0.32\textwidth]{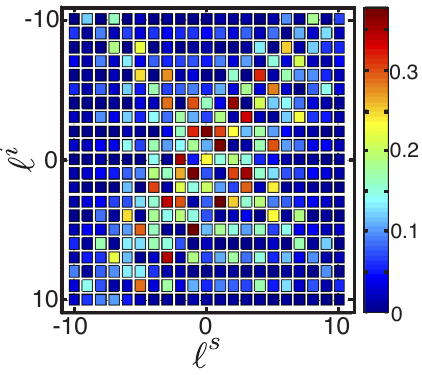}}\newline
\end{center}
\caption{Modulus of the transition element $F$, Eq.~(\ref{eq:Flslilp}),  as a function of the orbital angular momentum of the signal $\ell^s$ and idler $\ell^i$ photon. They involve a pump Bessel Gauss photon with transverse wave number $\kappa_\bot^p=0.01\mu$m$^{-1}$, a signal Bessel photon with $\kappa_\bot^s=0.0001\mu$m$^{-1}$ and an idler photon with $\kappa_\bot^i=0.01\mu$m$^{-1}$.   The idler and signal photons are emitted with their main propagation axis with orientation angles $\tilde\theta=\theta_{ec}$ and $\tilde\varphi_s = -\pi/2$ and $\tilde\varphi_i = \pi/2$
for the first row, while  $\tilde\varphi_s = 0$ and $\tilde\varphi_i = \pi$
in the second row. The pump angular quantum number is $\ell^p=0$ in figures (a) and (d), $\ell^p =1$ in figures (b) and (e), and $\ell^p = 2$ in figures (c) and (f). The BBO crystal length is 1mm, its optical axis  is located in the $Y$-$Z$ plane, and the width of the transversal wave number for the Bessel-Gauss photons is $W= 0.0005\mu$m$^{-1}$.  The optical axis of the crystal is located in the $Y$-$Z$ plane  $\mathbf{a} \sim (0,0.49,0.87).$}\label{F:01}
\end{figure*}

\begin{figure*}
\begin{center}
\subfloat[]{\label{F:L0-MPI-05}\includegraphics[width=0.32\textwidth]{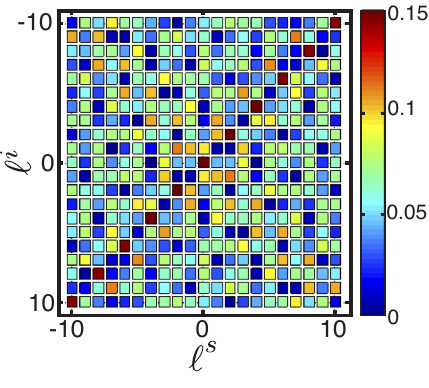}}
\subfloat[]{\label{F:L1-MPI-05}\includegraphics[width=0.32\textwidth]{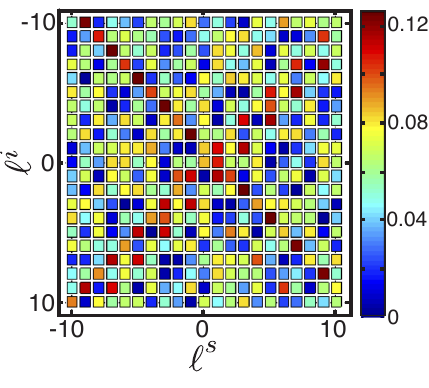}}
\subfloat[]{\label{F:L2-MPI-05}\includegraphics[width=0.32\textwidth]{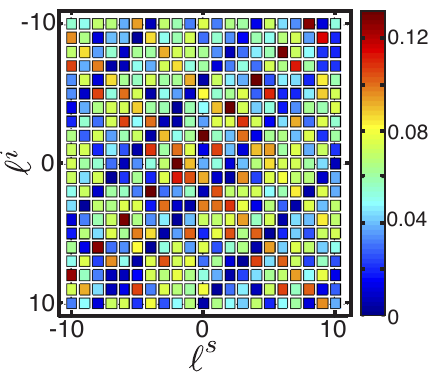}}\newline
\subfloat[]{\label{F:L0-0PI-05}\includegraphics[width=0.32\textwidth]{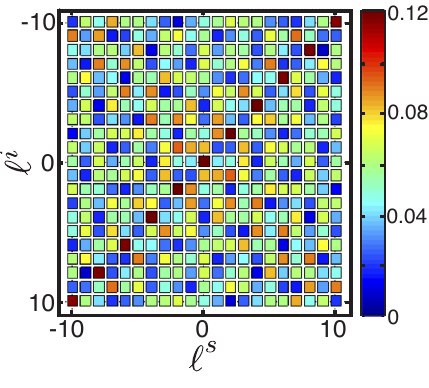}}
\subfloat[]{\label{F:L1-0PI-05}\includegraphics[width=0.32\textwidth]{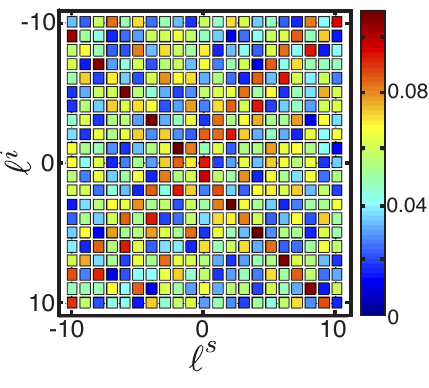}}
\subfloat[]{\label{F:L2-0PI-05}\includegraphics[width=0.32\textwidth]{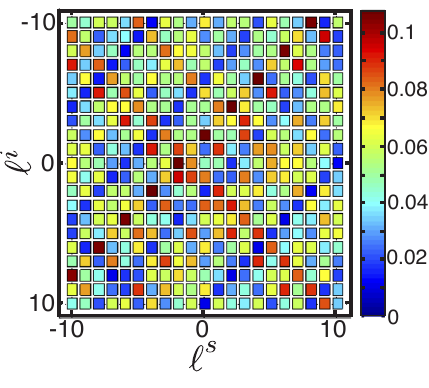}}\newline
\end{center}
\caption{Modulus of the transition element $F$, Eq.~(\ref{eq:Flslilp}),  as a function of the orbital angular momentum $\ell^s$ of the signal and $\ell^i$ of the idler photon. They involve a pump Bessel Gauss photon with transverse wave number $\kappa_\bot^p=0.05\mu$m$^{-1}$, a signal Bessel photon with $\kappa_\bot^s=0.001\mu$m$^{-1}$ and an idler photon with $\kappa_\bot^i=0.05\mu$m$^{-1}$. The idler and signal photons are emitted with their main propagation axis with orientation angles $\tilde\theta=\theta_{ec}$ and $\tilde\varphi_s = -\pi/2$ and $\tilde\varphi_i = \pi/2$
in the first row, while  $\tilde\varphi_s = 0$ and $\tilde\varphi_i = \pi$
in the second row, The pump angular quantum number is $\ell^p=0$ in figures (a) and (d), $\ell^p=1$ in figures (b) and (e) and $\ell^p= 2$ in figures (c) and (f).  The BBO crystal length is 1mm, its optical axis  is located in the $Y$-$Z$ plane and the width of the transversal wave number for the Bessel-Gauss photons is $W= 0.0005\mu$m$^{-1}$. The optical axis of the crystal is located in the $Y$-$Z$ plane  $\mathbf{a} \sim (0,0.49,0.87)$.}\label{F:05}
\end{figure*}

In Figs.~(\ref{F:01}) and (\ref{F:05}) we illustrate the behavior of the conditional
amplitudes as a function of the angular momentum of the pump photon as well as a function of its transverse wave vector.
In those figures we also illustrate the dependence on the orientation of the resulting idler and signal photons.
To make the results closer to expected experimental realizations, the calculations were performed with Bessel-Gauss
modes with a small waist  $W= 0.0005\mu$m$^{-1}$. The crystal properties are the same as in the illustrative examples
of the CAS and AS distributions in last section.

 As could be inferred from the results obtained for the CAS, smaller
values of the pump transverse vector $\kappa_\bot^p$ yield a more localized distribution of the OAM of the photon pair. That is, pump beams that can be closely described by the paraxial approximation can be used to generate photon pairs with few relevant values of $\ell^i$ and $\ell^s$. In particular, a paraxial pump beam with $\ell^p=0$ will yield mainly photon pairs without OAM along their propagation axis.

For pump beams
with bigger $\kappa_\bot^p$ it is predicted a high correlation between the angular momentum of the idler and signal photons involving a broad but well defined values $\ell^i$ and $\ell^s$ along a straight line. This property is preserved whether the structured photon pairs are detected in the direction of maximal CAS (first row in Figs.~(\ref{F:01}) and (\ref{F:05})) or in any other orientation as illustrated in the second row in Figs.~(\ref{F:01}) and (\ref{F:05}). For paraxial beams, the maximum transition rates for an orientation perpendicular to that of maximal CAS is approximately half of the maximum values found along the maximal CAS.

Notice that as the orbital angular momentum
is evaluated along different axes for the idler, signal and pump photons, it does not make strict sense to talk about  conservation of
OAM by comparing $\ell^i+\ell^s$ with $\ell^p$. Nevertheless we can observe that the greatest transition rates in the non paraxial
regime, Fig.~(\ref{F:05}), are along straight lines that go through the origin for $\ell^p$= 0, pass through $\vert\ell^i\vert$ = 1 and
$\ell^s = 0$ for $\ell^p$= 1, and   pass through $\vert\ell^i\vert$ =2 and $\ell^s = 0$ for $\ell^p$= 2. For photon pairs emitted in  an orientation perpendicular to that of maximal CAS this property is more evident. This could be a consequence of the fact that in this orientation the AS
is more symmetric under the change $\tilde\varphi^\prime = \tilde\varphi + \pi$, as illustrated in Fig.~1.

\begin{figure*}
\begin{center}
\subfloat[]{\label{MS:L0-MPI-01}\includegraphics[width=0.32\textwidth]{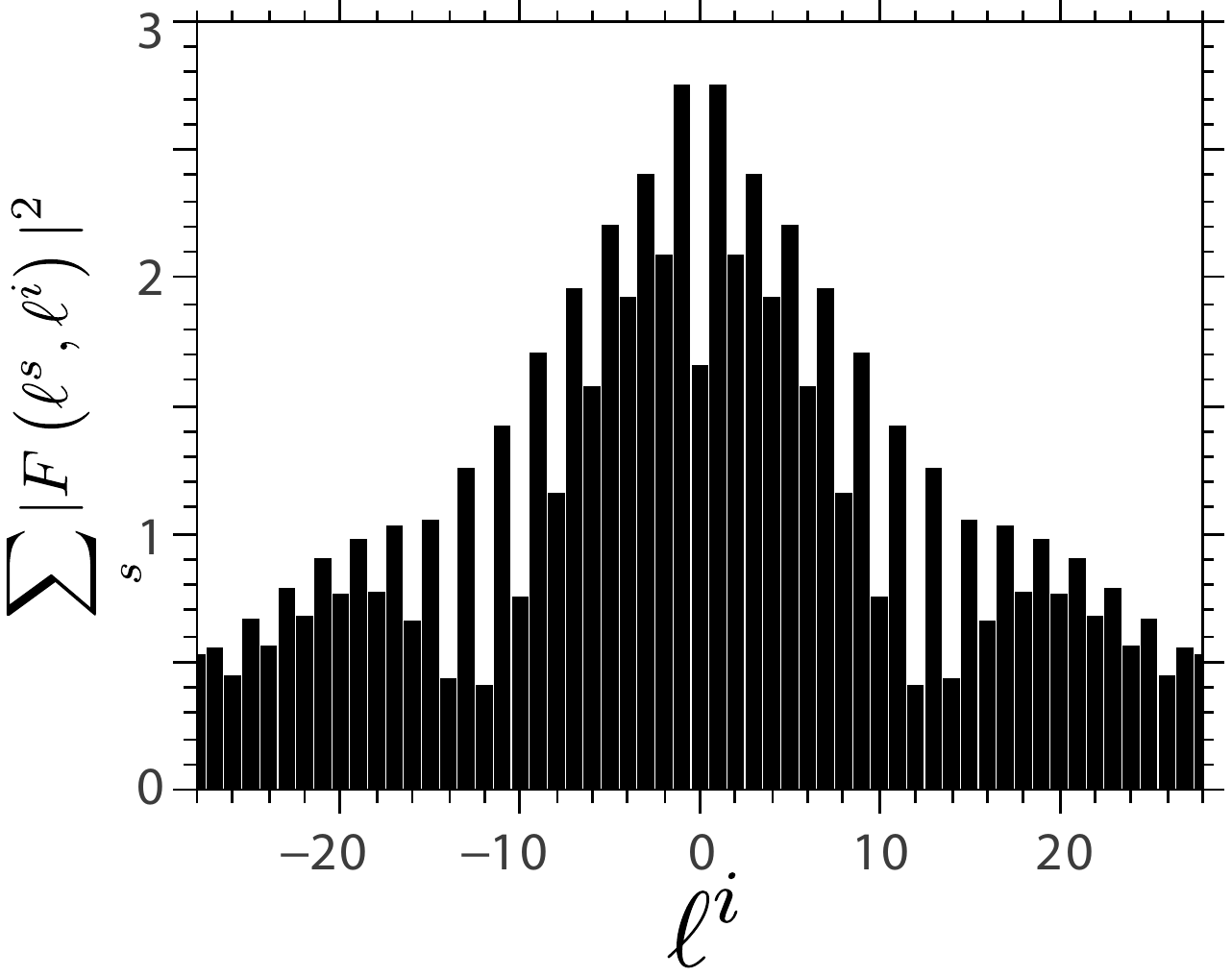}}
\subfloat[]{\label{MS:L1-MPI-01}\includegraphics[width=0.32\textwidth]{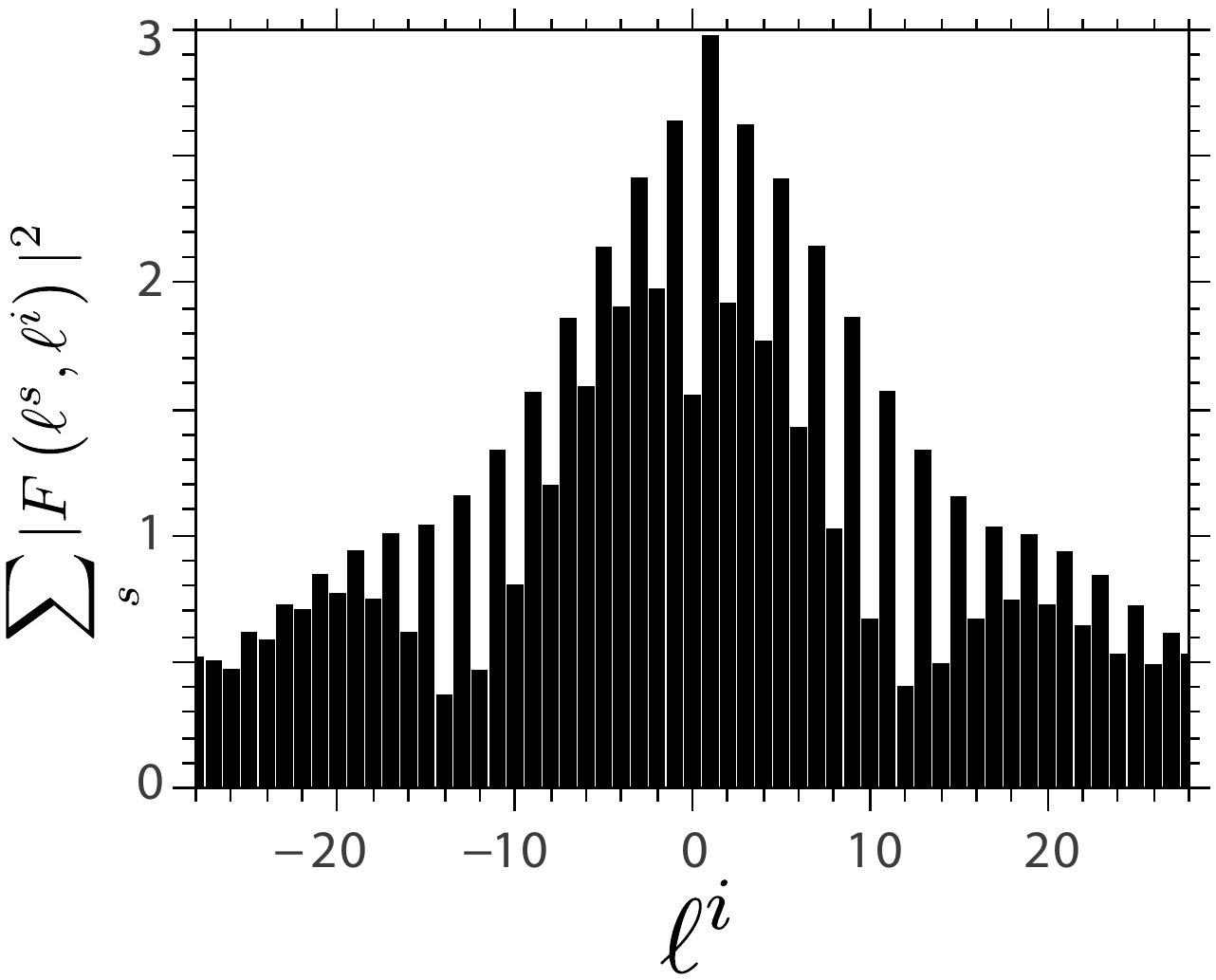}}
\subfloat[]{\label{MS:L2-MPI-01}\includegraphics[width=0.32\textwidth]{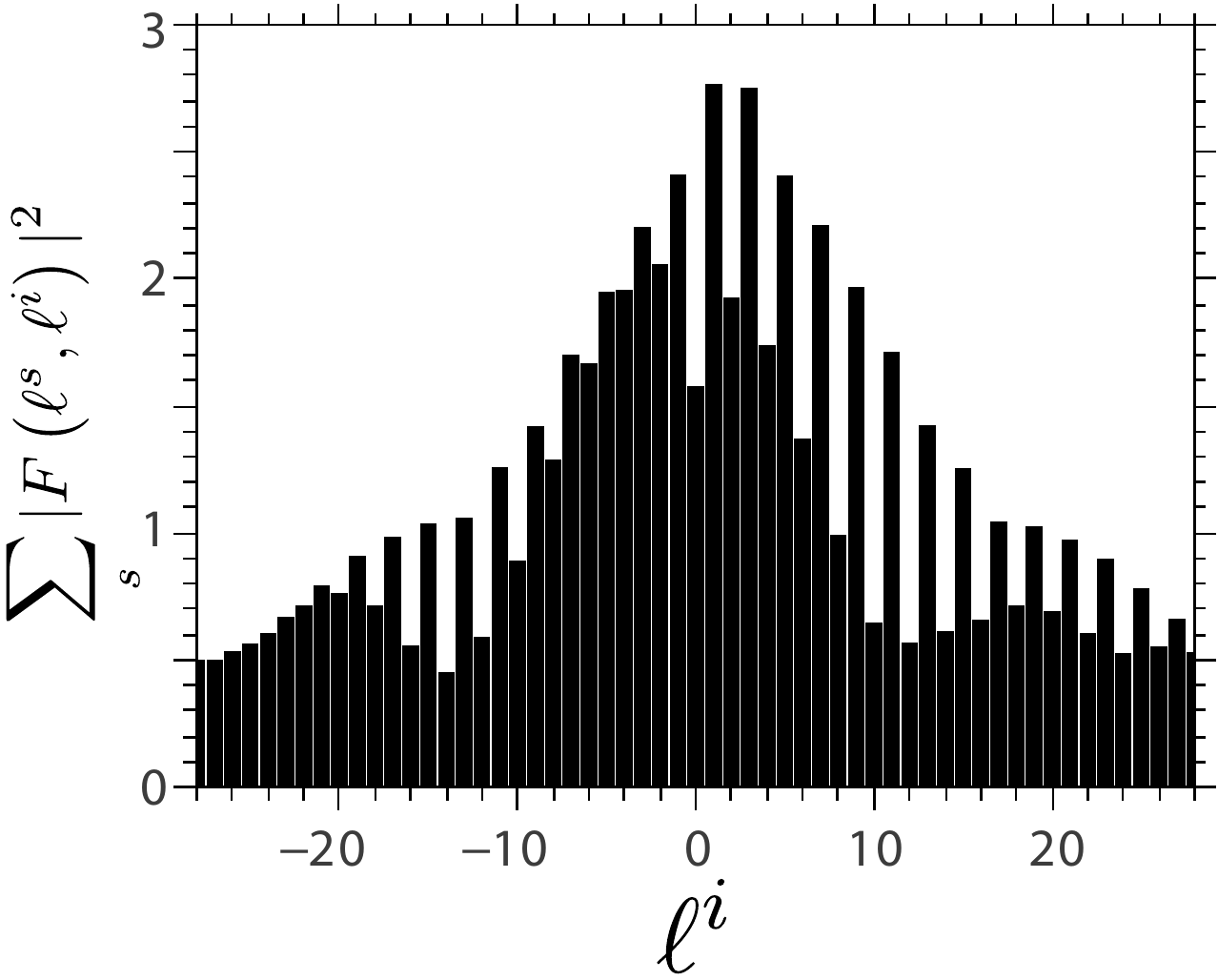}}\newline
\subfloat[]{\label{MI:L0-MPI-01}\includegraphics[width=0.32\textwidth]{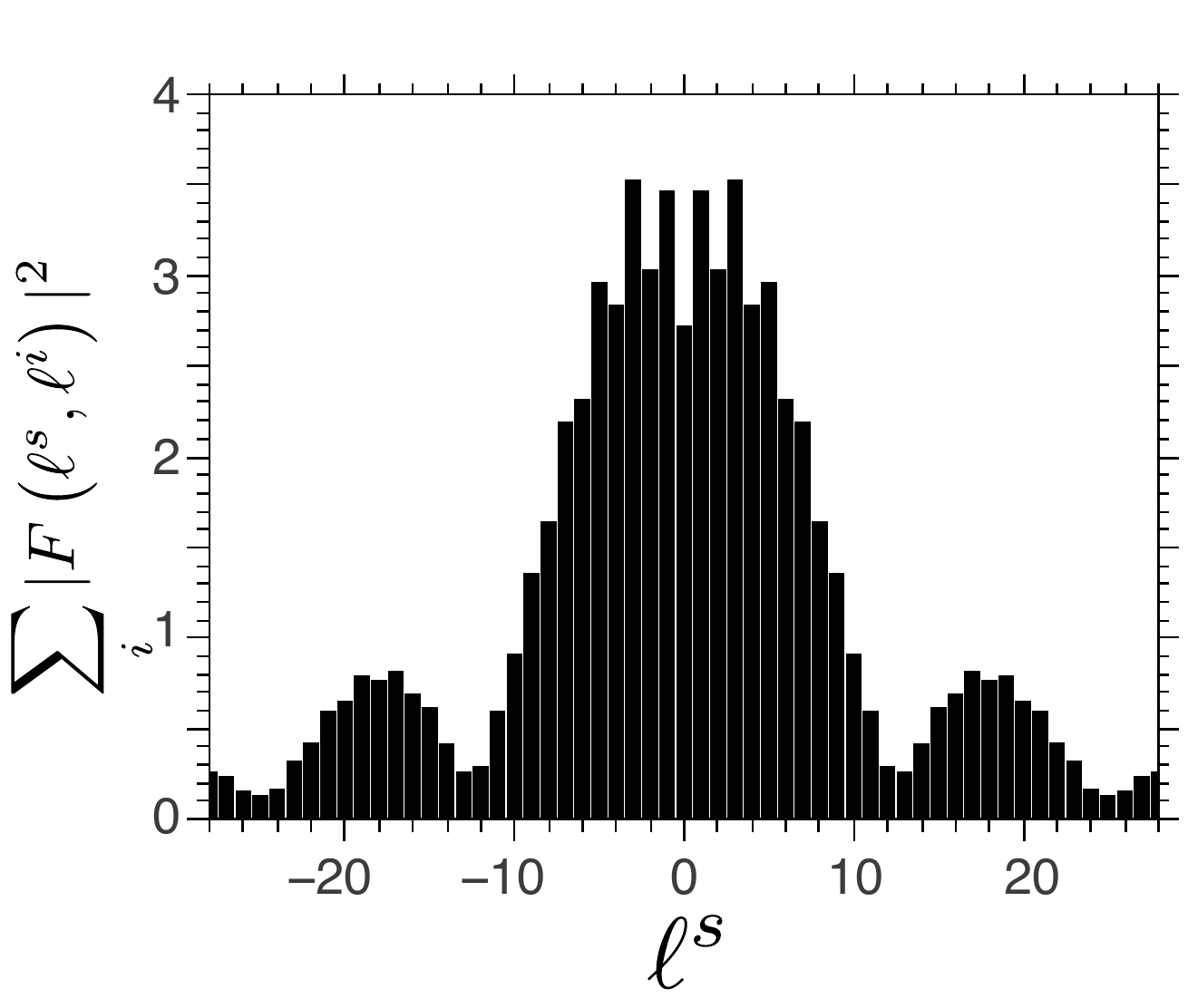}}
\subfloat[]{\label{MI:L1-MPI-01}\includegraphics[width=0.32\textwidth]{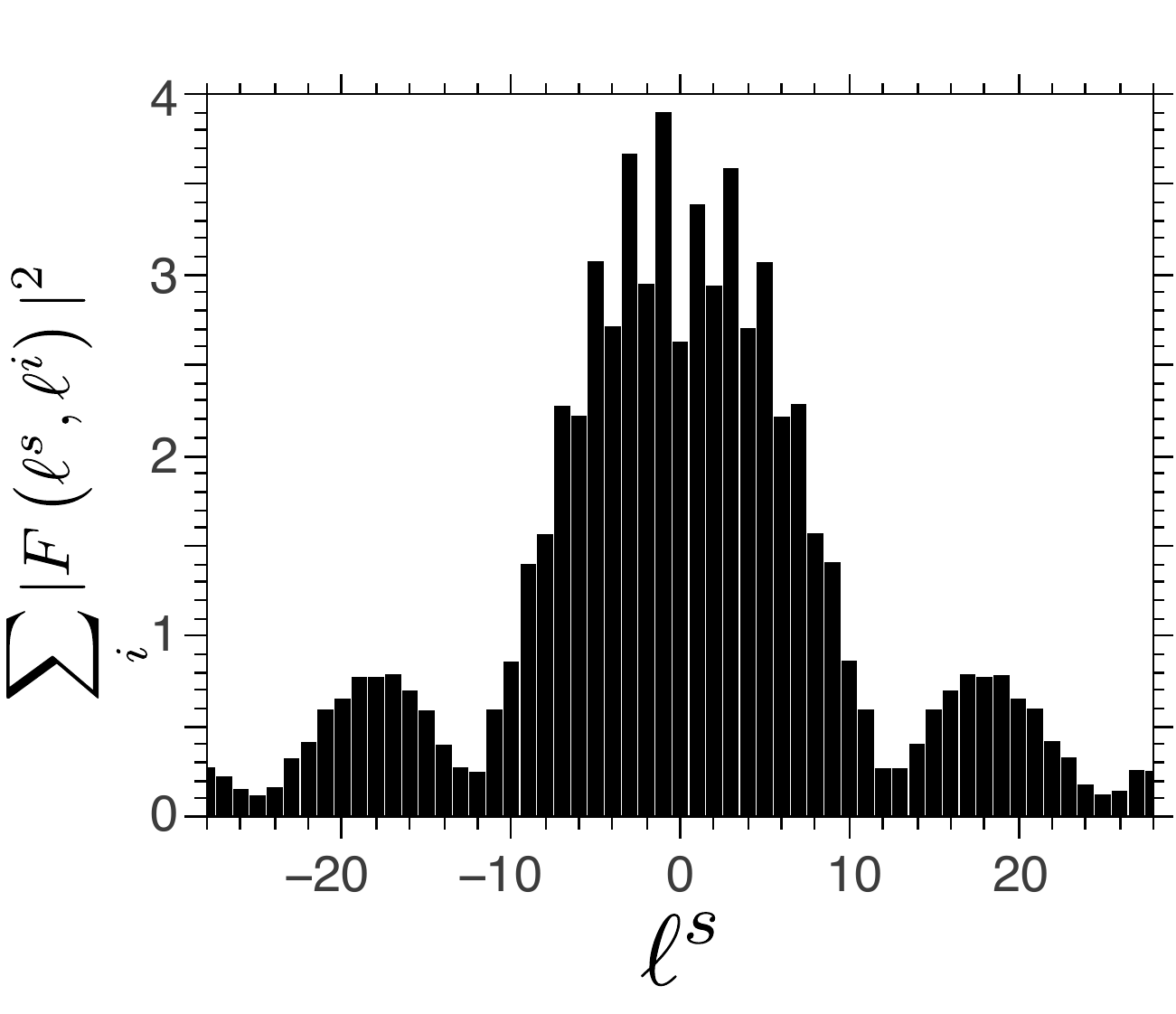}}
\subfloat[]{\label{MI:L2-MPI-01}\includegraphics[width=0.32\textwidth]{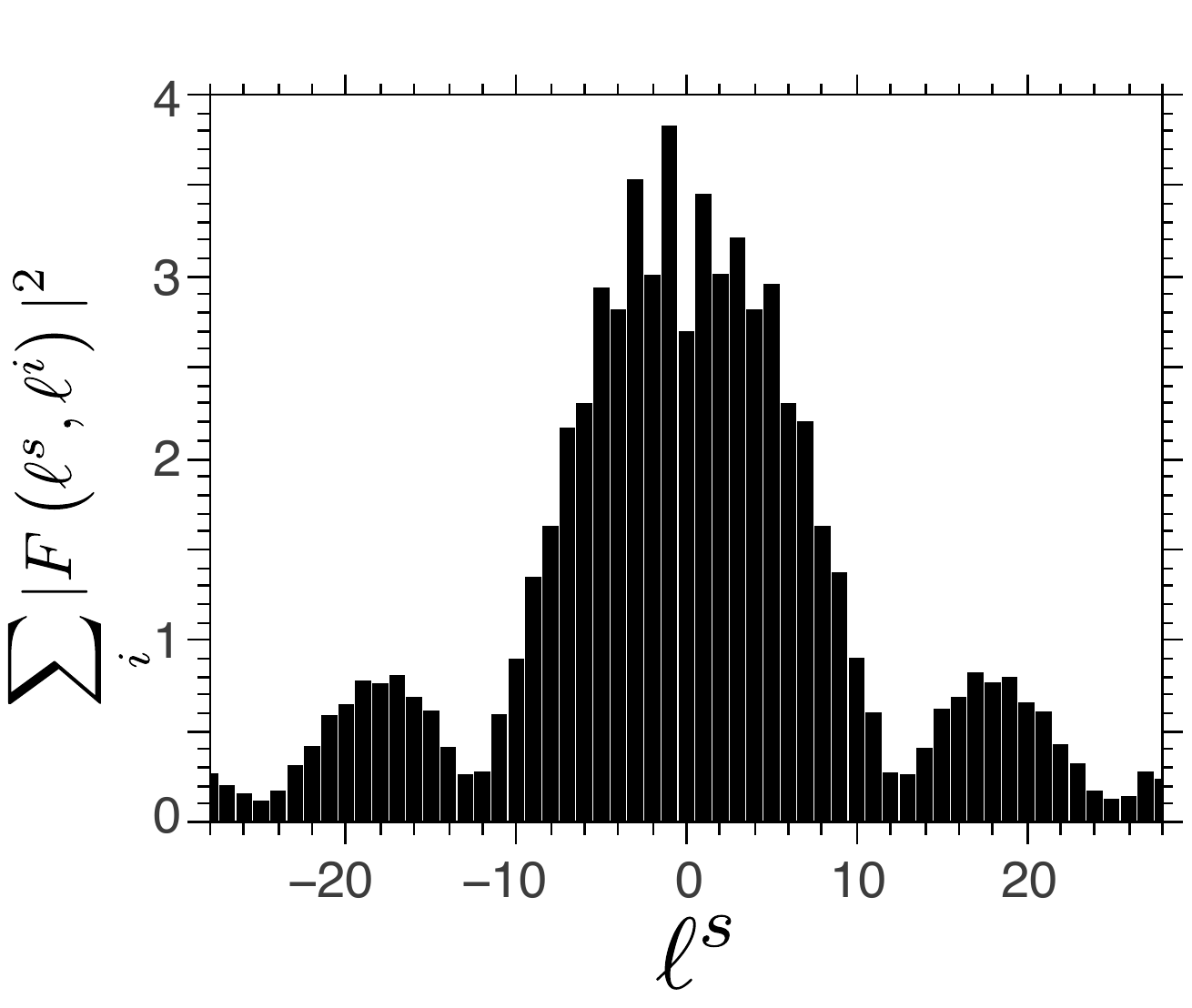}}\newline
\end{center}
\caption{Marginal distribution of orbital angular momentum of the signal $\ell^s$ and idler $\ell^i$ photon. They involve a pump Bessel Gauss profile with transverse wave number $\kappa_\bot^p=0.01\mu$m$^{-1}$, a signal Bessel photon with $\kappa_\bot^s=0.0001\mu$m$^{-1}$ and an idler photon with $\kappa_\bot^i=0.01\mu$m$^{-1}$. The idler and signal photons are emitted with their main propagation axis with orientation angles $\tilde\theta=\theta_{ec}$ and $\tilde\varphi_s = -\pi/2$ and $\tilde\varphi_i = \pi/2$
The pump angular quantum number is $\ell^p=0$ in figures (a) and (d), $\ell^p =1$ in figures (b) and (e), and $\ell^p = 2$ in figures (c) and (f).  The BBO crystal length is 1 mm, its optical axis  is located in the $Y$-$Z$ plane and the width of the transversal wave number for the Bessel-Gauss photons is $W= 0.0005\mu$m$^{-1}$. The optical axis of the crystal is located in the $Y$-$Z$ plane  $\mathbf{a} \sim (0,0.49,0.87).$}\label{F:marg}
\end{figure*}

Finally, in Fig.~(\ref{F:marg}), the marginal distributions of the idler and signal orbital angular momentum are illustrated. They were evaluated for a paraxial pump beam  with the same set up as in Fig.~(\ref{F:01}). For the photon in the pair with the same value of $\kappa_\bot$ than the pump beam, i. e., for idler photon, the OAM marginal distribution has clearly an oscillatory behavior dependent on the parity of the $\ell^i$ for $-15\le \ell^i\le 15$. Meanwhile, for the photon with lower value of $\kappa_\bot$ this behavior is observed for a much smaller $\ell^s$ interval.

\section{Conclusions}
In this paper, it has been shown that SPDC can be used to generate structured photons with several predetermined properties inherited from the pump beam. We have found that the use of scalar pump fields that are approximately propagation invariant can be used to generate heralded single photons with the same property.
A detailed study of type I spontaneous down conversion (SPDC) of Bessel-Gauss photons was performed as an interesting example for the study of generation of structured photons from beams that are propagation invariant and also have OAM. Explicit analytical and numerical results were given for the angular and conditional spectrum using a birefringent crystal cut with the optical axis optimized for SPDC with standard Gaussian pump beams. Since the AS and the CAS do not depend on the detailed phase structure of the pump beam in the wave vector domain, the results obtained in this work are expected to be valid for other structured pump beams such as  Mathieu and
Weber beams.

It was also shown that, as the magnitude of the transverse wave vector of a pump Bessel-Gauss beam increases, the emission cone is deformed into two non coaxial cones which touch each other along a line determined by the orientation of the optical axis of the nonlinear crystal. At this location, the conditional spectrum becomes maximal for a pair of photons, one of which is best described in wave vector space by a Gaussian like photon, with a very small transverse wave vector, and the other by a Bessel-Gauss photon with the same mean and width distribution of transverse wave vectors as the incident pump photon. Both of them have their main propagation axis close to the cone expected for a Gaussian pump beam. The results were successfully compared with reported experiments.

The study on the OAM distribution for the photon pairs showed the existence of clear correlations in the OAM quantum number of the idler and signal photons. These correlations can be manipulated by varying the pump parameters $\kappa_\bot^p$ and $\ell^p$. They also depend on the  orientation of the photon pair main propagation axes. The OAM significant correlations involve a smaller number of
$\ell^s$ and $\ell^i$ values for paraxial pump beams. This means that heralded values of $\ell^s$, given a particular value of $\ell^i$, could be easily obtained in that regime. However, even for relatively high values of $\kappa_\bot^p$, we observe some interesting OAM correlation features. In particular, there is a trend to preserve parity (the value of $\ell^s + \ell^i$ for the most probable photon pairs have the same parity that $\ell^p$). This effect results evident from the analysis of the marginal correlations. There is also a trend to get the greatest transition rates along straight lines in the $\ell^s$-- $\ell^i$ space. These lines pass, in general, through the $\ell^p$ value. Finally, it has also been shown that the OAM correlations cannot be interpreted in terms of angular momentum conservation.


\begin{thebibliography}{10}
\bibitem{fn} That is not the case for Laguerre-Gaussian  beams taken as pump beams in SPDC. They are paraxial beams that also carry orbital angular momentum; their angular spectra contains a factor $k_\bot^{n+2m}$ that yields a  dependence on $m$ of the corresponding joint amplitude
\end{thebibliography}
\end{document}